\documentclass[prc,amsmath,amssymb,superscriptaddress,showpacs,preprint,
tightenlines]{revtex4}
\usepackage{longtable,graphicx,dcolumn,epsfig}

\begin{document}

\title{Pygmy dipole strength in $^{90}$Zr}

\author{R.~Schwengner}
\affiliation{Institut f\"ur Strahlenphysik, Forschungszentrum
             Dresden-Rossendorf, D-01314 Dresden, Germany}
\author{G.~Rusev}
\altaffiliation{Present address: Department of Physics, Duke University, and
                Triangle Universities Nuclear Laboratory, Durham, NC 27708}
\affiliation{Institut f\"ur Strahlenphysik, Forschungszentrum
             Dresden-Rossendorf, D-01314 Dresden, Germany}
\author{N.~Tsoneva}
\affiliation{Institut f\"ur Theoretische Physik, Universit\"at Gie{\ss}en,
             D-35392 Gie{\ss}en, Germany}
\affiliation{Institute for Nuclear Research and Nuclear Energy, BAS,
             BG-1784 Sofia, Bulgaria}
\author{N.~Benouaret}
\altaffiliation{Permanent address: Facult\'e de physique, Universit\'e des
                Sciences et de la technologie d'Alger, El-Alia 16111,
                Bab-Ezzouar-Alger, Algerie}
\affiliation{Institut f\"ur Strahlenphysik, Forschungszentrum
             Dresden-Rossendorf, D-01314 Dresden, Germany}
\author{R.~Beyer}
\affiliation{Institut f\"ur Strahlenphysik, Forschungszentrum
             Dresden-Rossendorf, D-01314 Dresden, Germany}
\author{M.~Erhard}
\affiliation{Institut f\"ur Strahlenphysik, Forschungszentrum
             Dresden-Rossendorf, D-01314 Dresden, Germany}
\author{E.~Grosse}
\affiliation{Institut f\"ur Strahlenphysik, Forschungszentrum
             Dresden-Rossendorf, D-01314 Dresden, Germany}
\affiliation{Institut f\"ur Kern- und Teilchenphysik,
             Technische Universit\"at Dresden, D-01062 Dresden, Germany}
\author{A.~R.~Junghans}
\affiliation{Institut f\"ur Strahlenphysik, Forschungszentrum
             Dresden-Rossendorf, D-01314 Dresden, Germany}
\author{J.~Klug}
\altaffiliation{Present address: Ringhals Nuclear Power Plant,
                SE-43022 V\"ar\"obacka, Sweden}
\affiliation{Institut f\"ur Strahlenphysik, Forschungszentrum
             Dresden-Rossendorf, D-01314 Dresden, Germany}
\author{K.~Kosev}
\affiliation{Institut f\"ur Strahlenphysik, Forschungszentrum
             Dresden-Rossendorf, D-01314 Dresden, Germany}
\author{H.~Lenske}
\affiliation{Institut f\"ur Theoretische Physik, Universit\"at Gie{\ss}en,
             D-35392 Gie{\ss}en, Germany}
\author{C.~Nair}
\affiliation{Institut f\"ur Strahlenphysik, Forschungszentrum
             Dresden-Rossendorf, D-01314 Dresden, Germany}
\author{K.~D.~Schilling}
\affiliation{Institut f\"ur Strahlenphysik, Forschungszentrum
             Dresden-Rossendorf, D-01314 Dresden, Germany}
\author{A.~Wagner}
\affiliation{Institut f\"ur Strahlenphysik, Forschungszentrum
             Dresden-Rossendorf, D-01314 Dresden, Germany}

\date{\today}

\begin{abstract}
The dipole response of the $N=50$ nucleus $^{90}$Zr was studied in
photon-scattering experiments at the electron linear accelerator ELBE with
bremsstrahlung produced at kinetic electron energies of 7.9, 9.0, and 13.2 MeV.
We identified 189 levels up to an excitation energy of 12.9 MeV. Statistical
methods were applied to estimate intensities of inelastic transitions and to
correct the intensities of the ground-state transitions for their branching
ratios. In this way we derived the photoabsorption cross section up to the
neutron-separation energy. This cross section matches well the photoabsorption
cross section obtained from $(\gamma,n)$ data and thus provides information
about the extension of the dipole-strength distribution toward energies below
the neutron-separation energy. An enhancement of $E1$ strength has been found
in the range of 6 MeV to 11 MeV. Calculations within the framework of the
quasiparticle-phonon model ascribe this strength to a vibration of the
excessive neutrons against the $N = Z$ neutron-proton core, giving rise to a
pygmy dipole resonance.
\end{abstract}


\maketitle

\section{Introduction}
\label{sec:intro}

Gamma-ray strength functions, in particular dipole-strength functions, are an
important ingredient for the analysis of photodisintegration reactions as well
as of the inverse reactions like neutron capture. These reactions play an
important role for specific processes of the nucleosynthesis. Moreover, an
improved experimental and theoretical description of neutron-capture reactions
is important for next-generation nuclear technologies.

The measurement of dipole-strength distributions from low excitation energy up
to the neutron-separation energy delivers information about magnitude and
structure of the low-energy tail of the giant dipole resonance (GDR), in
particular about possible additional vibrational modes like the so-called pygmy
dipole resonance (PDR). Dipole-strength distributions up to neutron-separation
energies have been studied for only few nuclides in experiments with
monoenergetic photons (see, e.g., Refs. \cite{axe70,dat73,las87,ala87}) and in
experiments with bremsstrahlung (see, e.g., Ref. \cite{kne06} and Refs.
therein). The bremsstrahlung facility \cite{sch05} at the superconducting
electron accelerator ELBE of the research center Dresden-Rossendorf opens up
the possibility to study the dipole response of stable nuclei with even the
highest neutron-separation energies in photon-scattering experiments.
In the course of a systematic study of dipole-strength distributions for
varying neutron and proton numbers in nuclei around $A$ = 90 \cite{sch07,rus08}
we have investigated the $N$ = 50 nuclide $^{90}$Zr.

In earlier experiments, the 2$^+$ states at 2186, 3308, 3842, 4120, 4230,
4680, and the 1$^{(-)}$ states at 4580 and 5504 keV were studied with
bremsstrahlung produced by 5 MeV electrons \cite{met72,met74}. About 20 further
states were found in experiments at higher energies up to 10 MeV
\cite{can76,nds97}. For the 15 $J$ = 1 states above 6 MeV negative parity was
deduced from an experiment with polarized photons \cite{ber84}.

In the present study we identified about 190 levels up to 12.9 MeV. We applied
statistical methods to estimate the intensities of inelastic transitions to 
low-lying excited levels and to correct the intensities of the elastic
transitions to the ground state with their branching ratios. The
dipole-strength distribution deduced from the present experiments is compared
with predictions of the quasiparticle-phonon model.

\section{Experimental Methods and results}
\label{sec:exp}

The nuclide $^{90}$Zr was studied in photon-scattering experiments at the
superconducting electron accelerator ELBE of the research center
Dresden-Rossendorf. Bremsstrahlung was produced using electron beams of 7.9,
9.0, and 13.2 MeV kinetic energy. The average currents were about 330 $\mu$A
in the measurement at 7.9 MeV and about 500 $\mu$A in the measurements at
higher energy. The electron beams hit radiators consisting of niobium foils
with thicknesses of 4 $\mu$m in the low-energy measurements and of 7 $\mu$m
in the measurement at 13.2 MeV. A 10 cm thick aluminum absorber was placed
behind the radiator to reduce the low-energy part of the bremsstrahlung
spectrum (beam hardener). The collimated photon beam impinged onto the target
with a flux of about $10^9$ s$^{-1}$ in a spot of 38 mm diameter. The target
was a disk with a diameter of about 20 mm to enable an irradiation with a
constant flux density over the target area. The target consisted of 4054.2 mg
of $^{90}$ZrO$_2$ enriched to 97.7\%, combined with 339.5 mg of $^{11}$B,
enriched to 99.52\% and also formed to a disk of 20 mm diameter, that was used
for the determination of the photon flux.
Scattered photons were measured with four high-purity germanium (HPGe)
detectors of 100\% efficiency relative to a NaI detector of 3 in. diameter and
3 in. length. All HPGe detectors were surrounded by escape-suppression shields
made of bismuth germanate scintillation detectors. Two HPGe detectors were
placed vertically at 90$^\circ$ relative to the photon-beam direction at a
distance of 28 cm from the target. The other two HPGe detectors were positioned
in a horizontal plane at 127$^\circ$ to the beam at a distance of 32 cm from
the target. Absorbers of 8 mm Pb plus 3 mm Cu and of 3 mm Pb plus 3 mm Cu were
placed in front of the detectors at 90$^\circ$ and 127$^\circ$, respectively,
in the measurements at 7.9 and 9.0 MeV whereas in the measurement at 13.2 MeV
absorbers of 13 mm Pb plus 3 mm Cu and of 8 mm Pb plus 3 mm Cu were used for
the detectors at 90$^\circ$ and 127$^\circ$, respectively. A detailed
description of the bremsstrahlung facility is given in Ref.~\cite{sch05}.

\begin{figure}
\epsfig{file=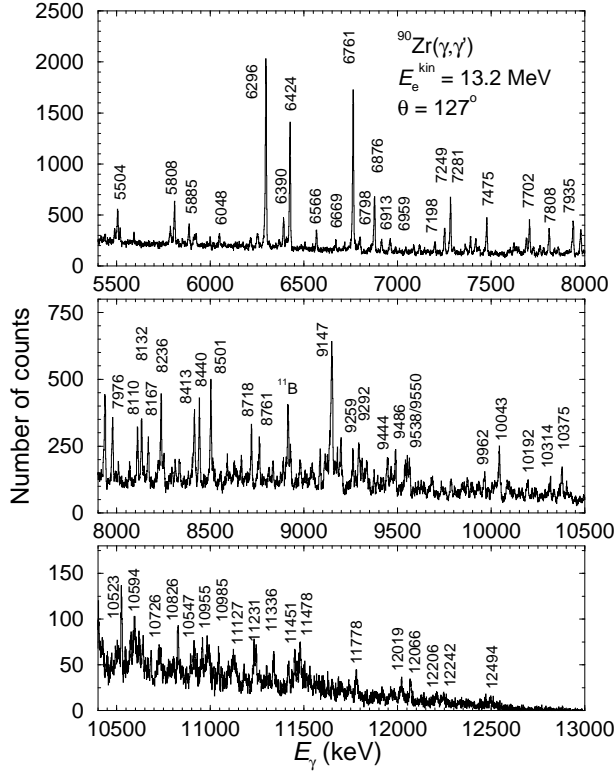,width=8cm}
\caption{\label{fig:spec}Parts of a spectrum of photons scattered from
$^{90}$Zr combined with $^{11}$B during the irradiation with bremsstrahlung
produced by electrons of an energy of $E_e^{\rm kin}$ = 13.2 MeV. This
spectrum is the sum of the spectra measured with the two detectors placed at
127$^\circ$ relative to the beam. The most dominant transitions assigned to
$^{90}$Zr are marked with their energies in keV.}
\end{figure}

Spectra of scattered photons were measured for 56 h, 65 h, and 97 h in the
experiments at 7.9, 9.0, and 13.2 MeV electron energy, respectively. Parts of
a spectrum including events measured with the two detectors placed at
127$^\circ$ relative to the beam at an electron energy of 13.2 MeV are shown in
Fig.~\ref{fig:spec}.

\subsection{The photon-scattering method}
\label{sec:meth}

In photon-scattering experiments the energy-integrated scattering cross section
$I_s$ of an excited state at the energy $E_x$ can be deduced from the measured
intensity of the respective transition to the ground state (elastic
scattering). It can be determined relative to the known integrated scattering
cross sections $I_s(E_x^{\rm B})$ of states in $^{11}$B \cite{ajz90}:

\begin{equation}
\label{eq:sigs}
\frac{I_s(E_x)}{I_s(E_x^{\rm B})} =
\left(\frac{I_\gamma(E_\gamma,\theta)} {W(E_\gamma,\theta)
\Phi_\gamma(E_x) N_N}\right)
\left(\frac{I_\gamma(E_\gamma^{\rm B},\theta)}
{W(E_\gamma^{\rm B},\theta) \Phi_\gamma(E_x^{\rm B}) N_N^{\rm B}}\right)^{-1}.
\end{equation}

Here, $I_\gamma(E_\gamma,\theta)$ and $I_\gamma(E_\gamma^{\rm B},\theta)$
denote the measured intensities of a considered ground-state transition at
$E_\gamma$ and of a ground-state transition in $^{11}$B at $E_\gamma^{\rm B}$,
respectively, observed at an angle $\theta$ to the beam. $W(E_\gamma,\theta)$
and $W(E_\gamma^{\rm B},\theta)$ describe the angular correlations of these
transitions. The quantities $N_N$ and $N_N^{\rm B}$ are the numbers of
nuclei in the $^{90}$Zr and $^{11}$B targets, respectively. The quantities
$\Phi_\gamma(E_x)$ and $\Phi_\gamma(E_x^{\rm B})$ stand for the photon fluxes
at the energy of the considered level and at the energy of a level in $^{11}$B,
respectively.

The integrated scattering cross section is related to the partial width of the
ground-state transition $\Gamma_0$ according to
\begin{equation}
\label{eq:gam}
I_s = \int \sigma_{\gamma \gamma} ~dE =
            \left(\frac{\pi \hbar c}{E_x}\right)^2
            \frac{2 J_x + 1}{2 J_0 + 1}
            \frac{\Gamma_0^2}{\Gamma},
\end{equation}
where $\sigma_{\gamma \gamma}$ is the elastic scattering cross section, $E_x$,
$J_x$ and $\Gamma$ denote energy, spin and total width of the excited level,
respectively, and $J_0$ is the spin of the ground state.

For the determination of the level widths one is faced with two problems.
First, a considered level can be fed by transitions from higher-lying states.
The measured intensity of the ground-state transition is in this case higher
than the one resulting from a direct excitation only. As a consequence, the
integrated cross section deduced from this intensity contains a part
originating from feeding in addition to the true integrated scattering cross
section: $I_{s+f} = I_s + I_f$. Furthermore, a considered level can deexcite
not only to the ground state, but also to low-lying excited states
(inelastic scattering). In this case, not all observed $\gamma$ transitions
are ground-state transitions. To deduce the partial width of a ground-state
transition $\Gamma_0$ and the integrated absorption cross section one has to
know the branching ratio $b_0 = \Gamma_0 / \Gamma$. If this branching ratio
cannot be determined, only the quantity $\Gamma_0^2 / \Gamma$ can be deduced
(cf. Eq.~(\ref{eq:gam})).

Spins of excited states can be deduced by comparing experimental ratios of
intensities measured at two angles with theoretical predictions. The optimum
combination are angles of 90$^\circ$ and 127$^\circ$ because the respective
ratios for the spin sequences $0-1-0$ and $0-2-0$ differ most at these
angles. The expected values are $W(90^\circ)/W(127^\circ)_{0-1-0} = 0.74$
and $W(90^\circ)/W(127^\circ)_{0-2-0} = 2.18$ taking into account opening
angles of $16^\circ$ and $14^\circ$ of the detectors at $90^\circ$ and
$127^\circ$, respectively.

\subsection{Detector response and photon flux}
\label{sec:effflux}

For the determination of the integrated scattering cross sections according
to Eq. (\ref{eq:sigs}) the relative efficiencies of the detectors and the
relative photon flux are needed. For the determination of the dipole-strength
distribution described in Sec. \ref{sec:corr} the experimental spectrum has to
be corrected for detector response, for the absolute efficiency, and for the
absolute photon flux, for background radiation, and for atomic processes
induced by the impinging photons in the target material. The detector response
has been simulated using the program package GEANT3 \cite{cer93}. The
reliability of the simulation was tested by comparing simulated spectra with
measured ones as described in Refs.~\cite{sch07,rus08}.

The absolute efficiencies of the HPGe detectors were determined experimentally
up to 1333 keV from measurements with $^{22}$Na, $^{60}$Co, $^{133}$Ba, and
$^{137}$Cs calibration sources. The efficiency curve calculated with GEANT3
was scaled to fit the absolute experimental values. To check the simulated
efficiency curve at higher energies we used an uncalibrated $^{56}$Co source
produced in-house using the $^{56}$Fe$(p,n)^{56}$Co reaction. The relative
efficiency values obtained from the $^{56}$Co source adjusted to the calculated
value at 1238 keV are consistent with the simulated curve as is illustrated
in Ref. \cite{sch07}.

\begin{figure}
\epsfig{file=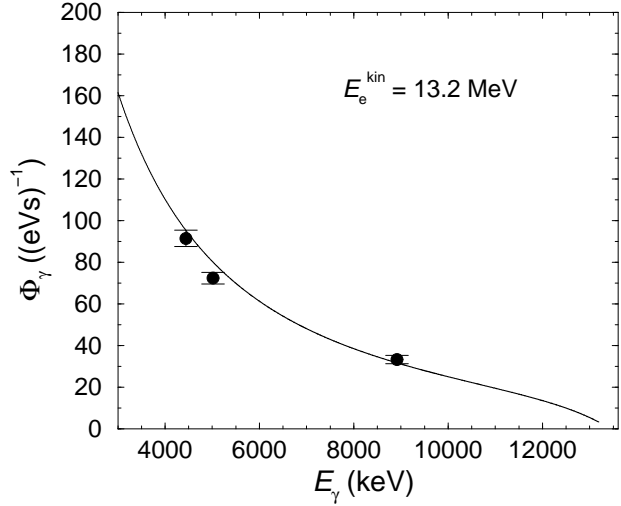,width=8cm}
\caption{\label{fig:flux}Absolute photon flux at the target deduced from
intensities of transitions in $^{11}$B (circles) using the calculated
efficiency and the relative photon flux calculated as described in the text
(solid line).}
\end{figure}

The absolute photon flux was determined from intensities and known
integrated scattering cross sections of transitions in $^{11}$B which was
combined with the $^{90}$Zr target (cf. Sec. \ref{sec:exp}). For interpolation,
the photon flux was calculated using a code \cite{hau05} based on the
approximation given in Ref.~\cite{roc72} and including a screening correction
according to Ref.~\cite{sal87}. In addition, the flux was corrected for the
attenuation by the beam hardener. This flux was adjusted to the experimental
values as is shown in Fig.~\ref{fig:flux}. The transition from the level at
7285 keV could not be used because of superposition by another transition.

\subsection{Experiments at various electron energies}
\label{sec:ene}

\begin{figure}
\epsfig{file=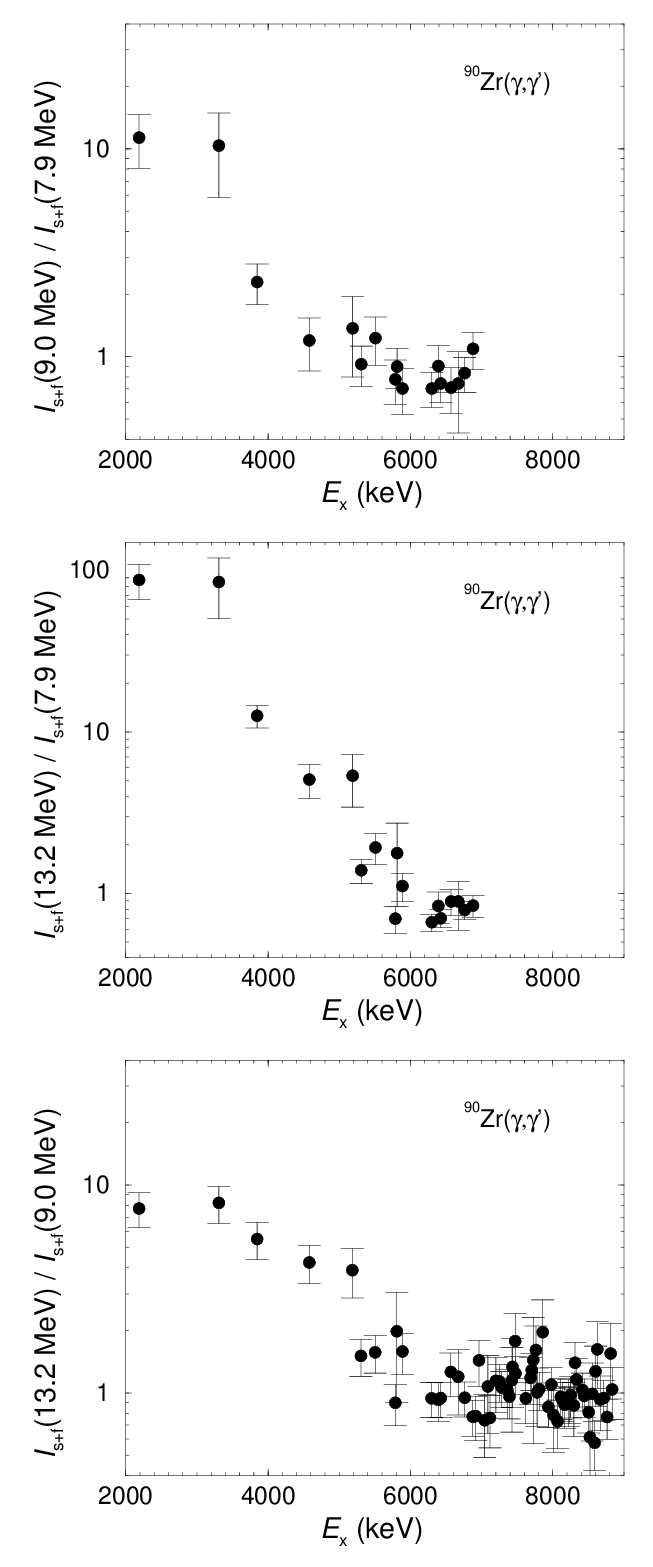,width=8cm}
\caption{\label{fig:sigs}Ratios of integrated cross sections $I_{s+f}$ of
transitions in $^{90}$Zr obtained at different electron energies.}
\end{figure}

Measurements at various electron energies allow us to estimate the influence of
feeding on the integrated cross sections. Ratios of the quantities $I_{s+f}$
obtained for levels in $^{90}$Zr from measurements at different electron
energies are shown in Fig.~\ref{fig:sigs}.
The plotted ratios reveal that (i) only levels below $E_x \approx$ 6 MeV are
influenced considerably by feeding and (ii) levels in the range of
$E_x \approx$ 4 to 6 MeV are mainly fed by levels above $E_x \approx$ 9 MeV.
Transitions found in the measurement at $E_e^{\rm kin}$ = 7.9 MeV are assumed
to be ground-state transitions. Transitions additionally observed up to 7.9 MeV
in the measurements at 9.0 MeV and 13.2 MeV are therefore considered as
inelastic transitions from high-lying to low-lying excited states. In the same
way, transitions in the range from 7 -- 9 MeV observed in the measurement at
9.0 MeV are considered as ground-state transitions, whereas transitions in this
energy range found additionally at 13.2 MeV are considered as branchings to
low-lying states. By comparing the measurements in this way, transitions from
high-lying to excited low-lying states may be filtered out. The remaining
transitions, assumed as ground-state transitions, have been used to derive the
corresponding level energies which are listed in Table~\ref{tab:gam} together
with spin assignments deduced from angular distributions of the ground-state
transitions, integrated scattering cross sections, and the quantities
$\Gamma_0^2 / \Gamma$.

\begin{longtable*}{rrrrrrrr}
\caption{\label{tab:gam}Levels assigned to $^{90}$Zr.} \\
\hline\hline
$E_x$ (keV) \footnotemark[1] &
$\frac{I_\gamma(90^\circ)}{I_\gamma(127^\circ)}$\footnotemark[2] &
$J_x^\pi$\footnotemark[3] &
$\frac{I_{s+f}( 9.0)}{I_{s+f}(7.9)}$\footnotemark[4] &
$\frac{I_{s+f}(13.2)}{I_{s+f}(7.9)}$\footnotemark[4] &
$\frac{I_{s+f}(13.2)}{I_{s+f}(9.0)}$\footnotemark[4] &
$I_s$ (eV b) \footnotemark[5]  &
$\Gamma_0^2/\Gamma$ (meV) \footnotemark[6] \\
\hline
 2186.2(1) &         & 2$^+$\footnotemark[7] &11.3(33) &88(22)   & 7.7(15) & 19(4)   &   \\
 3308.0(2) &         & 2$^+$\footnotemark[7] &10.4(45) &85(35)   & 8.2(17) &  4.7(19)&   \\
 3842.0(2) &         & 2$^+$\footnotemark[7] &  2.3(5) &12.6(20) & 5.5(11) & 26(4)   &   \\
 3932.4(6) &         &       &         &         &         &  8.3(32)&  \\
 4507.0(8) &         &       &         &         &         & 21(10)  &  \\
 4578.3(3) &         &       & 1.2(3)  & 5.1(12) & 4.2(9)  & 16(4)   &  \\
 5183.0(5) &         &       & 1.4(6)  & 5.4(19) & 3.9(10) &  7.1(24)&  \\
 5304.5(3) &         &       & 0.92(20)& 1.4(2)  & 1.5(3)  & 57(8)   &  \\
 5503.6(3) &         &       & 1.2(3)  & 1.9(4)  & 1.6(3)  & 34(6)   &  \\
 5785.0(4) &         &       & 0.78(19)& 0.70(13)& 0.90(20)& 50(8)   & 145(22)\footnotemark[8]  \\
 5807.9(3) &         &       & 0.90(20)& 1.8(9)  & 2.0(11) & 78(10)  & 228(30)\footnotemark[8]  \\
 5884.4(4) &         &       & 0.70(17)& 1.1(2)  & 1.6(4)  & 48(8)   & 143(23)\footnotemark[8]  \\
 6295.8(2) & 0.75(3) & 1$^-$\footnotemark[7] & 0.70(13)& 0.66(8) & 0.94(18)&740(63)  &2545(218) \\
 6389.8(3) & 0.82(8) & 1     & 0.90(23)& 0.84(19)& 0.93(20)& 82(15)  & 290(54)  \\
 6424.3(2) & 0.74(4) & 1$^-$\footnotemark[7] & 0.74(14)& 0.70(9) & 0.94(18)&479(43)  &1716(153) \\
 6565.7(3) & 0.79(15)& 1     & 0.71(18)& 0.89(17)& 1.3(3)  & 66(9)   & 245(34)  \\
 6669.2(7) & 0.64(16)& 1     & 0.74(31)& 0.89(30)& 1.2(4)  & 29(8)   & 111(32)  \\
 6761.4(2) & 0.76(3) & 1$^-$\footnotemark[7] & 0.83(16)& 0.79(10)& 0.95(18)&644(60)  &2553(237) \\
 6875.4(2) & 0.73(4) & 1$^-$ & 1.09(22)& 0.84(13)& 0.77(15)&198(22)  & 813(91)  \\
 6960.4(7) & 0.77(11)& 1     &         &         & 1.4(36) & 44(10)  & 183(41)  \\
 7042.0(7) & 0.74(19)& 1     &         &         & 0.74(25)& 25(6)   & 109(28)  \\
 7085.6(10)& 0.83(21)& (1)   &         &         & 1.1(4)  & 30(10)  & 130(43)  \\
 7198.2(6) & 0.49(20)& 1     &         &         & 1.1(3)  & 45(10)  & 201(47)  \\
 7249.0(3) & 0.78(7) & 1$^-$\footnotemark[7] &         &         & 1.1(2)  & 99(17)  & 450(79)  \\
 7280.9(7)\footnotemark[9] &         &       &         &         &         &         &          \\
 7361.0(6) & 0.57(17)& 1     &         &         & 1.0(3)  & 33(7)   & 153(33)  \\
 7387.6(4) & 0.68(10)& 1     &         &         & 0.96(21)& 75(14)  & 355(65)  \\
 7424.5(10)& 1.4(6)  &       &         &         & 1.1(5)  & 14(5)   &  69(24)\footnotemark[8]\\
 7433.8(8) & 0.52(21)& 1     &         &         & 1.3(5)  & 19(6)   &  93(28)  \\
 7468(2)   & 1.5(5)  &       &         &         & 1.8(6)  & 12(4)   &  61(18)\footnotemark[8]  \\
 7474.9(3) & 0.96(10)& (1)   &         &         & 1.2(3)  &127(23)  & 617(113) \\
 7685.8(4) & 0.65(10)& 1     &         &         & 1.2(3)  & 70(13)  & 358(68)  \\
 7702.9(3) & 0.83(7) & 1$^-$\footnotemark[7] &         &         & 1.3(3)  &158(28)  & 815(142) \\
 7723.1(9) & 0.9(3)  &       &         &         & 1.4(9)  & 20(6)   & 106(31)\footnotemark[8]  \\
 7759.7(6) & 0.82(19)& (1)   &         &         & 1.6(5)  & 38(9)   & 198(45)  \\
 7779.0(6) & 0.77(17)& 1     &         &         & 1.0(3)  & 40(9)   & 212(47)  \\
 7807.9(3) & 0.75(6) & 1     &         &         & 1.0(2)  &125(23)  & 659(120) \\
 7857.8(7) & 0.84(25)& (1)   &         &         & 2.0(8)  & 34(9)   & 183(50)  \\
 7935.6(3) & 0.81(6) & 1     &         &         & 0.86(17)&209(36)  &1141(199) \\
 7976.6(4) & 0.67(7) & 1     &         &         & 1.1(2)  &125(23)  & 691(124) \\
 8006.9(8) & 0.71(25)& 1     &         &         & 0.78(27)& 36(9)   & 197(49)  \\
 8067.4(5) & 0.99(27)& (1)   &         &         & 0.73(19)& 55(13)  & 312(74)  \\
 8110.0(8) & 0.71(9) & 1$^-$\footnotemark[7] &         &         & 0.96(21)&124(24)  & 704(136) \\
 8131.9(4) & 0.74(9) &(1$^-$)\footnotemark[7]&         &         & 0.93(20)&154(28)  & 884(164) \\
 8144(2)   &                 &         &         &         & 20(13)  & 115(75)\footnotemark[8]  \\ 
 8166.7(5) & 0.89(14)& (1)   &         &         & 0.88(20)& 98(19)  & 566(111) \\
 8221.2(8) & 0.80(13)& 1     &         &         & 0.92(24)& 57(12)  & 332(73)  \\
 8235.6(3) & 0.74(4) & 1     &         &         & 0.89(18)&254(44)  &1495(260) \\
 8250.7(5) & 0.68(7) & 1     &         &         & 0.98(22)& 85(16)  & 504(98)  \\
 8295.3(10)& 0.89(29)& (1)   &         &         & 0.87(25)& 40(11)  & 241(67)  \\
 8313.0(7) & 0.73(18)& 1     &         &         & 1.4(4)  & 70(15)  & 420(95)  \\
 8334.1(5) & 0.72(18)& 1     &         &         & 1.2(3)  & 90(20)  & 539(121) \\
 8357.5(18)& 0.32(18)& 1     &         &         &         & 16(7)   &  97(42)  \\
 8382.1(10)& 1.1(4)  & (1)   &         &         &         & 25(6)   & 157(32)  \\
 8403.7(11)&         &       &         &         &         & 43(7)   & 263(43)\footnotemark[8]  \\ 
 8413.5(4) & 0.87(9) & 1     &         &         & 1.0(2)  &212(38)  &1300(235) \\
 8440.6(4) & 0.87(10)& 1     &         &         & 0.96(20)&224(40)  &1382(250) \\
 8467.7(15)& 1.0(5)  &       &         &         &         & 31(17)  & 193(106) \\
 8501.2(4) & 0.81(8) & 1$^-$\footnotemark[7] &         &         & 0.81(17)&346(63)  &2166(393) \\
 8518(3)   & 1.6(6)  &       &         &         & 0.61(33)& 40(16)  & 253(98)\footnotemark[8]  \\
 8544(4)   &         &       &         &         &         &  8(3)   &  51(19)\footnotemark[8]  \\
 8553.5(12)& 0.12(6) & 1     &         &         &         & 79(8)   & 504(50)  \\
 8588.3(7) & 0.58(13)& 1     &         &         & 0.57(15)& 93(21)  & 597(133) \\
 8598.2(10)& 0.59(22)& 1     &         &         & 1.3(4)  & 42(12)  & 270(78)  \\
 8625.6(10)& 0.72(27)& 1     &         &         & 1.6(6)  & 37(11)  & 239(69)  \\
 8664.1(5) & 0.68(16)& 1     &         &         & 0.93(27)& 59(14)  & 385(92)  \\
 8716.6(5) & 0.75(6) & 1$^-$\footnotemark[7] &         &         & 0.94(20)&176(33)  &1162(220) \\
 8751.0(8) & 0.36(14)& 1     &         &         &         & 62(15)  & 410(97)  \\
 8760.4(5) & 0.58(10)& 1     &         &         & 0.77(17)&162(31)  &1077(203) \\
 8812.0(13)& 0.76(24)& 1     &         &         & 1.5(6)  & 37(13)  & 246(86)  \\
 8833.2(8) & 0.67(15)& 1     &         &         & 1.0(3)  & 83(20)  & 561(134) \\
 8874.9(9) & 0.37(13)& 1     &         &         &         & 41(11)  & 280(75)  \\
 8903.0(8) &         &       &         &         &         & 57(6)   & 394(43)\footnotemark[8]  \\
 8927.4(4) &         &       &         &         &         &127(13)  & 878(91)\footnotemark[8]  \\
 8978.4(9) & 0.8(4)  & (1)   &         &         &         & 88(31)  & 615(215) \\
 8985(2)   &         &       &         &         &         & 45(13)  & 315(90)\footnotemark[8]  \\
 9004.7(5) & 0.45(23)& 1     &         &         &         & 34(11)  & 239(78)  \\
 9014.0(8) &         &       &         &         &         & 24(14)  & 170(101)\footnotemark[8] \\
 9034.0(8) &         &       &         &         &         & 35(7)   & 247(50)\footnotemark[8]  \\
 9043.6(4) & 0.44(6) & 1     &         &         &         & 71(10)  & 503(68)  \\
 9053.5(7) &         &       &         &         &         & 38(7)   & 270(49)\footnotemark[8]  \\
 9085.1(3) & 0.60(9) & 1     &         &         &         &129(15)  & 925(109) \\
 9111.1(6) & 0.64(15)& 1     &         &         &         &141(20)  &1018(140) \\
 9123.6(7) &         &       &         &         &         &126(17)  & 913(126)\footnotemark[8] \\
 9137.5(7) &         &       &         &         &         &185(22)  &1338(163)\footnotemark[8] \\
 9148.5(3) & 0.60(4) & 1$^-$\footnotemark[7] &         &         &         &703(66)  &5103(480) \\
 9164.9(7) &         &       &         &         &         &107(14)  & 778(104)\footnotemark[8] \\
 9177.5(5) &         &       &         &         &         &162(18)  &1182(134)\footnotemark[8] \\
 9187(3)   &         &       &         &         &         & 45(13)  & 329(95)\footnotemark[8]  \\
 9196.5(3) & 0.72(7) &(1$^-$)\footnotemark[7]&         &         &         &252(25)  &1849(186) \\
 9260.5(6) & 0.67(9) & 1     &         &         &         &149(19)  &1109(140) \\
 9292.8(5) & 0.65(7) & 1     &         &         &         &216(24)  &1619(181) \\
 9309.4(7) & 0.71(10)& 1     &         &         &         &137(18)  &1033(135) \\
 9333.4(6) & 0.67(10)& 1$^-$\footnotemark[7] &         &         &         &141(19)  &1064(146) \\
 9373.2(7) &         &       &         &         &         &111(21)  & 843(161)\footnotemark[8]\\
 9392.4(8) & 0.45(13)& 1     &         &         &         &102(19)  & 780(145) \\
 9409.4(11)&         &       &         &         &         & 71(16)  & 543(123)\footnotemark[8]\\
 9424.3(10)&         &       &         &         &         & 79(17)  & 608(129)\footnotemark[8]\\
 9444.7(4) & 0.65(8) & 1     &         &         &         &221(28)  &1707(219) \\
 9465.1(5) & 0.66(11)& 1     &         &         &         &169(25)  &1317(193) \\
 9486.8(4) & 0.75(10)& 1     &         &         &         &226(32)  &1765(251) \\
 9510.5(13)& 0.88(35)& (1)   &         &         &         & 45(16)  & 350(124) \\
 9524.1(13)& 0.47(22)& 1     &         &         &         & 44(14)  & 348(112) \\
 9539.2(5) & 0.85(13)& 1     &         &         &         &154(22)  &1213(175) \\
 9551.4(6) & 0.47(9) & 1     &         &         &         &160(23)  &1267(185) \\
 9563.0(6) & 0.64(17)& 1     &         &         &         &180(28)  &1424(221) \\
 9609.2(7) &         &       &         &         &         & 72(22)  & 573(177)\footnotemark[8]\\
 9625.1(8) &         &       &         &         &         & 58(16)  & 469(129)\footnotemark[8]\\
 9640.4(8) & 0.6(3)  & 1     &         &         &         & 56(14)  & 456(115) \\
 9666.0(8) & 0.7(3)  & (1)   &         &         &         & 39(9)   & 318(71)  \\
 9678.3(7) & 0.67(19)&(1$^-$)\footnotemark[7]&         &         &         & 67(10)  & 545(83)  \\
 9686.9(6) & 0.66(17)& 1     &         &         &         & 72(11)  & 591(89)  \\
 9733.2(5) & 0.5(3)  & 1     &         &         &         & 55(8)   & 450(68)  \\
 9741.7(7) &         &       &         &         &         & 33(6)   & 275(51)\footnotemark[8]\\
 9754.0(6) & 0.46(29)& 1     &         &         &         & 50(9)   & 413(71)  \\
 9784.6(5) & 1.37(27)&       &         &         &         &111(15)  & 921(122)\footnotemark[8]\\
 9805.4(10)& 1.6(6)  &       &         &         &         & 41(9)   & 341(73)\footnotemark[8]\\
 9843.4(6) & 0.77(16)& 1     &         &         &         & 84(15)  & 702(126) \\
 9855.5(8) & 0.55(17)& 1     &         &         &         & 58(13)  & 488(109) \\
 9872.4(4) & 0.49(12)& 1     &         &         &         &126(24)  &1067(207) \\
 9890.7(13)& 0.7(5)  & (1)   &         &         &         & 81(34)  & 686(289) \\
 9901.9(13)&         &       &         &         &         & 70(28)  & 592(242)\footnotemark[8]\\
 9932.1(12)& 0.43(14)& 1     &         &         &         &123(39)  &1053(330) \\
 9962.8(5) & 0.57(14)& 1     &         &         &         &172(37)  &1479(322) \\
 9984.1(11)&         &       &         &         &         & 69(34)  & 593(297)\footnotemark[8]\\
10004.2(10)& 0.42(19)& 1     &         &         &         & 70(16)  & 606(137) \\
10019.6(11)& 0.65(16)& 1     &         &         &         & 94(16)  & 819(138) \\
10031(2)   &         &       &         &         &         & 69(16)  & 599(136)\footnotemark[8]\\
10042.9(4) & 0.70(8) &(1$^-$)\footnotemark[7]&         &         &         &316(36)  &2762(310) \\
10083.8(6) & 0.78(13)& 1     &         &         &         & 93(13)  & 823(118) \\
10094.2(7) & 0.74(19)& 1     &         &         &         & 83(14)  & 732(120) \\
10104.9(12)& 0.7(3)  & (1)   &         &         &         & 49(13)  & 433(118) \\
10123.7(18)& 0.5(3)  & 1     &         &         &         &137(100) &1219(886) \\
10146.8(9) & 0.57(27)& 1     &         &         &         & 46(13)  & 409(114) \\
10163.4(8) & 0.50(23)& 1     &         &         &         & 60(16)  & 540(147) \\
10193.0(5) & 0.60(12)& 1     &         &         &         &168(25)  &1514(220) \\
10216.8(10)& 0.57(24)& 1     &         &         &         & 76(18)  & 693(159) \\
10233(4)   &         &       &         &         &         & 47(38)  & 427(348)\footnotemark[8] \\
10241(2)   & 0.9(3)  & (1)   &         &         &         & 79(34)  & 723(305) \\
10260.9(11)&         &       &         &         &         & 23(6)   & 212(53)\footnotemark[8] \\
10270.0(7) &         &       &         &         &         & 34(9)   & 308(81)\footnotemark[8]\\
10286.2(6) & 0.64(19)& 1     &         &         &         & 42(9)   & 388(79)  \\
10298.3(10)& 0.8(3)  & (1)   &         &         &         & 32(8)   & 290(69)  \\
10306.6(9) & 0.71(21)& 1     &         &         &         & 46(9)   & 426(80)  \\
10315.1(4) & 0.63(12)& 1     &         &         &         &103(14)  & 952(126) \\
10334.9(6) & 0.68(21)& 1     &         &         &         & 51(10)  & 470(93)  \\
10361(2)   & 0.86(29)& (1)   &         &         &         & 54(14)  & 504(129) \\
10376.8(4) & 0.46(7) & 1     &         &         &         &240(28)  &2240(259) \\
10402.5(9) & 0.72(14)& 1     &         &         &         & 85(16)  & 799(150) \\
10494.5(11)& 0.7(3)  & (1)   &         &         &         & 43(10)  & 411(92)  \\
10507.9(8) & 0.6(3)  & 1     &         &         &         & 49(10)  & 467(96)  \\
10524.6(4) & 0.72(10)& 1     &         &         &         &143(18)  &1376(175) \\
10595.0(7) & 0.80(19)& 1     &         &         &         & 92(14)  & 901(136) \\
10618.7(8) & 0.74(23)& 1     &         &         &         & 67(12)  & 651(119) \\
10638.5(9) & 0.41(20)& 1     &         &         &         & 59(12)  & 575(122) \\
10682.2(6) & 0.54(16)& 1     &         &         &         & 42(10)  & 417(95)  \\
10713.2(12)& 0.8(4)  & (1)   &         &         &         & 37(20)  & 369(195) \\
10728.2(11)& 0.67(20)& 1     &         &         &         &102(32)  &1022(316) \\
10827.1(5) & 0.82(15)& 1     &         &         &         &105(16)  &1068(167) \\
10914(2)   & 0.79(21)& (1)   &         &         &         &113(21)  &1168(214) \\
10957(2)   & 0.35(8) & 1     &         &         &         &118(19)  &1224(202) \\
10987.0(10)& 0.66(13)& 1     &         &         &         &161(23)  &1691(242) \\
11044(2)   &         &       &         &         &         & 49(17)  & 522(179)\footnotemark[8]\\
11094.2(15)&         &       &         &         &         & 70(10)  & 743(106)\footnotemark[8]\\
11108.0(16)&         &       &         &         &         & 39(8)   & 422(84)\footnotemark[8]\\
11120.4(9) & 0.60(16)& 1     &         &         &         & 92(17)  & 992(179) \\
11129.2(17)&         &       &         &         &         & 57(18)  & 611(196)\footnotemark[8]\\
11140(2)   &         &       &         &         &         & 57(9)   & 612(95)\footnotemark[8]\\
11232.4(7) & 0.45(14)& 1     &         &         &         & 88(13)  & 963(146) \\
11243.2(6) & 0.58(15)& 1     &         &         &         & 92(14)  &1010(152) \\
11337.7(6) & 0.85(13)& 1     &         &         &         & 91(15)  &1010(167) \\
11417.5(7) & 0.8(4)  & (1)   &         &         &         &108(25)  &1217(287) \\
11452.2(10)& 0.42(12)& 1     &         &         &         &132(28)  &1501(316) \\
11479.7(8) & 0.58(15)& 1     &         &         &         &191(33)  &2181(374) \\
11501(3)   &         &       &         &         &         & 66(37)  & 756(425)\footnotemark[8]\\
11510(7)   &         &       &         &         &         & 33(15)  & 382(172)\footnotemark[8]\\
11531(2)   & 0.42(25)& 1     &         &         &         & 74(30)  & 848(346) \\
11627.9(9) &         &       &         &         &         & 44(16)  & 518(182)\footnotemark[8]\\
11651.5(8) & 1.0(3)  & (1)   &         &         &         & 48(16)  & 564(185) \\
11777.4(10)& 0.46(22)& 1     &         &         &         &124(40)  &1489(479) \\
11788(3)   & 0.44(29)& 1     &         &         &         & 73(36)  & 885(440) \\
11963.3(18)& 0.8(3)  & (1)   &         &         &         & 68(14)  & 845(179) \\
11984(2)   & 0.57(18)& 1     &         &         &         & 57(13)  & 715(167) \\
12020.6(8) & 0.67(14)& 1     &         &         &         &155(21)  &1942(260) \\
12067.8(9) & 0.63(17)& 1     &         &         &         &124(19)  &1570(237) \\
12208.3(12)& 0.40(15)& 1     &         &         &         & 72(16)  & 929(214) \\
12243.6(14)& 0.57(19)& 1     &         &         &         & 62(15)  & 799(188) \\
12496.3(18)&         &       &         &         &         & 87(18)  &1181(244)\footnotemark[8]\\
12880.3(10)&         &       &         &         &         & 11(3)   & 164(46)\footnotemark[8]\\
\hline\hline
\footnotetext[1]{
Excitation energy. The uncertainty in parentheses is given in units of the last
digit. This energy was deduced from the $\gamma$-ray energy measured at
127$^\circ$ to the beam by including a recoil and Doppler correction.}
\footnotetext[2]{
Ratio of the intensities of the ground-state transitions measured at angles of
90$^\circ$ and 127$^\circ$. The expected values for an elastic pure-dipole
transition (spin sequence $0-1-0$) and for an elastic quadrupole transition
(spin sequence $0-2-0$) are 0.74 and 2.18, respectively.}
\footnotetext[3]{
Spin and parity of the state.}
\footnotetext[4]{
Ratio of integrated scattering+feeding cross sections deduced at different
electron energies. The deviation from unity is a measure of feeding.}
\footnotetext[5]{
Integrated scattering cross section. The values up to the level at 6875 keV
were deduced from the measurement at $E_e^{\rm kin}$ = 7.9 MeV, those from the
level at 6960 keV up to the level at 8832 keV were deduced from the measurement
at $E_e^{\rm kin}$ = 9.0 MeV, and the values given for levels at higher
energies were deduced from the measurement at $E_e^{\rm kin}$ = 13.2 MeV.}
\footnotetext[6]{
Partial width of the ground-state transition $\Gamma_0$ multiplied with the
branching ratio $b_0 = \Gamma_0/\Gamma$. The values deduced from the given
$I_s$ for states up to the 5504 keV state exceed the values of previous studies
\cite{nds97} outside their uncertainties which may indicate that the present
values are influenced by feeding. These values are therefore not given in this
table. An estimate of the partial width $\Gamma_0$ can be obtained from
$\Gamma_0^2/\Gamma$ after correction for a mean branching ratio
$b_0 = \Gamma_0/\Gamma$ = 0.9(1)\% for resolved ground-state transitions
(cf. Sec.~\ref{sec:corr}).}
\footnotetext[7]{
Spin and parity have been known from previous work (cf. Sec. \ref{sec:intro}).}
\footnotetext[8]{
Value deduced under the assumption of $J_x$ = 1.}
\footnotetext[9]{
This transition is superimposed by a transition in $^{11}$B.}
\end{longtable*}

In general, the complete level scheme may be constructed by searching for
combinations of $\gamma$-ray energies that fit another $\gamma$-ray according
to the Ritz principle. However, if one includes all observed transitions one
finds a huge number of possible two-fold or even three-fold combinations
\cite{sch07}.
Because we cannot derive complete information about the many possible branching
transitions, we are not able to determine the branching ratios of the
ground-state transitions from the present data. Therefore, we applied
statistical methods to estimate the intensities of branching transitions and
thus to correct the dipole-strength distribution.

\section{Determination of the dipole-strength distribution}
\label{sec:corr}

In the following we describe the statistical methods used to estimate the
intensity distribution of branching transitions and the branching ratios of the
ground-state transitions.

\begin{figure}
\epsfig{file=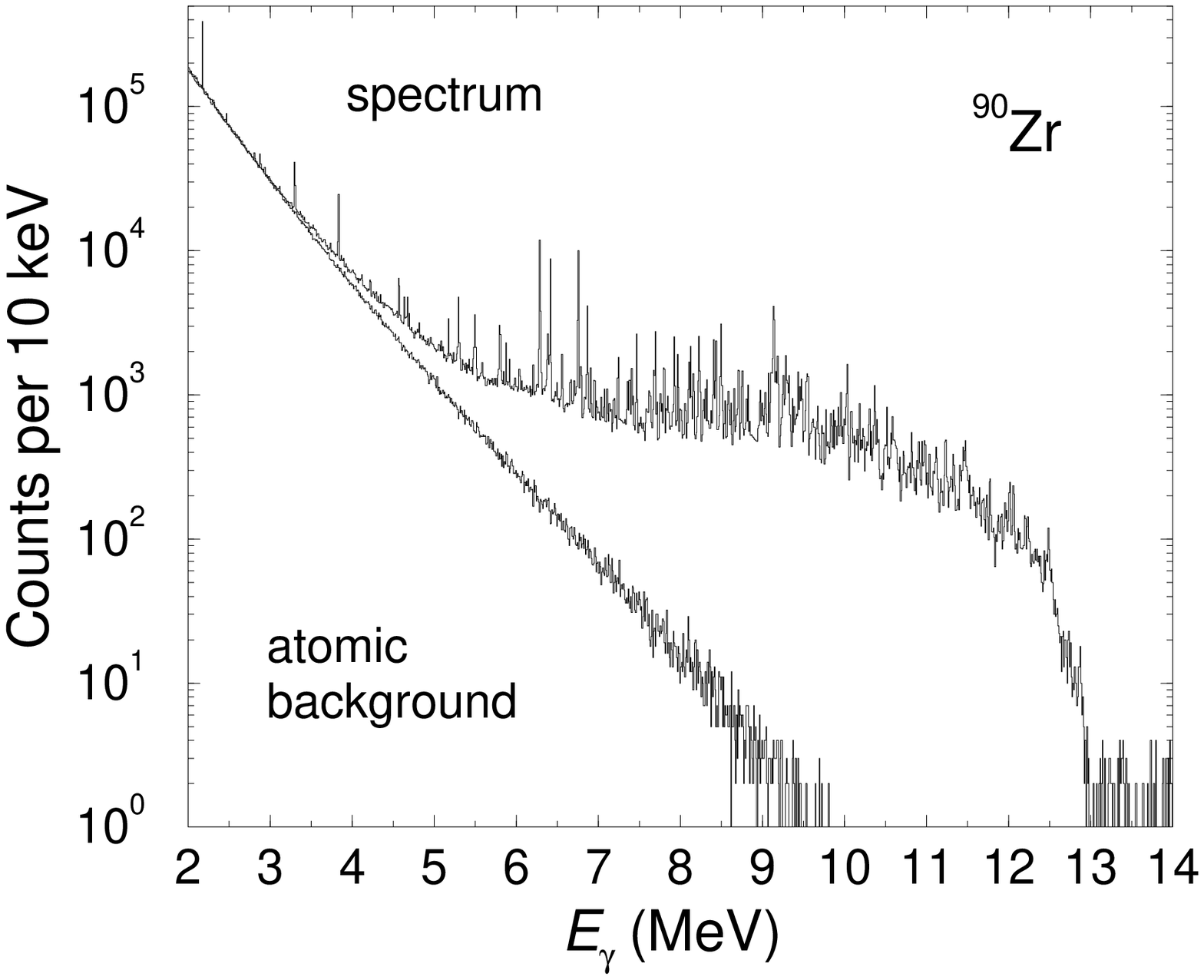,width=8cm}
\caption{\label{fig:cont}Experimental spectrum of $^{90}$Zr (corrected for room
background and detector response) and simulated spectrum of atomic background
(multiplied with efficiency and measuring time).}
\end{figure}

By using simulations as described in Sec. \ref{sec:effflux} we corrected the
spectrum including the two detectors at 127$^\circ$, measured during the
irradiation of the $^{90}$Zr target at $E_e^{\rm kin}$ = 13.2 MeV. In a first
step spectra of the ambient background adjusted to the intensities of the
1460.5 keV transition (decay of $^{40}$K) and 2614.9 keV transition
(decay of $^{208}$Tl) in the in-beam spectrum were subtracted from the measured
spectrum. It turned out that transitions following $(n,\gamma)$ reactions in
the HPGe detectors and in surrounding materials are negligibly small and thus,
did not require correction. To correct the spectrum for detector response,
spectra of monoenergetic $\gamma$ rays were calculated in steps of 10 keV by
using GEANT3. Starting from the high-energy end of the experimental spectrum,
the simulated spectra were subtracted sequentially. The resulting spectrum
including the two detectors at 127$^\circ$ is shown in Fig.~\ref{fig:cont}.

The background produced by atomic processes in the $^{90}$Zr target was
obtained from a GEANT3 simulation using the absolute photon flux deduced from
the intensities of the transitions in $^{11}$B (cf. Fig.~\ref{fig:flux}). The
corresponding background spectrum multiplied with the efficiency curve and with
the measuring time is also depicted in Fig.~\ref{fig:cont}.
As can be seen in Fig.~\ref{fig:cont} the continuum in the spectrum of photons
scattered from $^{90}$Zr is clearly higher than the background by atomic
scattering. This continuum is formed by a large number of non-resolvable
transitions with small intensities which are a consequence of the increasing
nuclear level density at high energy and of Porter-Thomas fluctuations of the
decay widths \cite{por56} in connection with the finite detector resolution
(e.g. $\Delta E \approx$ 7 keV at $E_\gamma \approx$ 9 MeV).

\begin{figure}
\epsfig{file=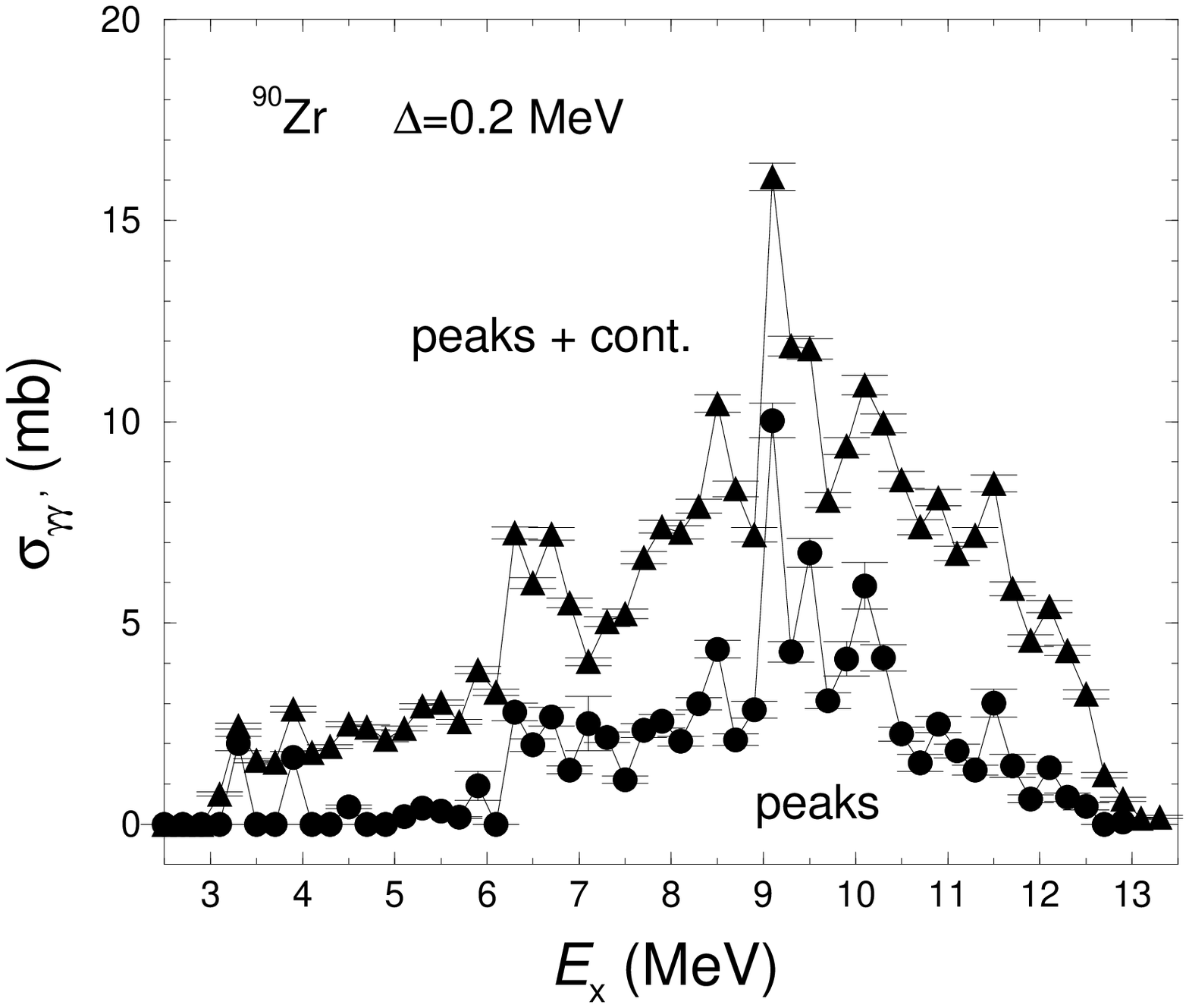,width=8cm}
\caption{\label{fig:siggg} Scattering cross sections in
$^{90}$Zr, derived as $\sigma_{\gamma \gamma'} = \sum_\Delta I_s/\Delta$, not
corrected for branching and averaged over energy bins of $\Delta$ = 0.2 MeV,
as derived from the difference of the experimental spectrum and the atomic
background shown in Fig.~\ref{fig:cont} (``peaks + cont.'', triangles) and
from the resolved peaks only (``peaks'', circles).}
\end{figure}

The relevant intensity of the photons resonantly scattered from $^{90}$Zr is
obtained from a subtraction of the atomic background from the
response-corrected experimental spectrum. The remaining intensity distribution
includes the intensity contained in the resolved peaks as well as the intensity
of the ``nuclear'' continuum. The scattering cross sections
$\sigma_{\gamma \gamma'}$ derived from this intensity distribution for energy
bins of 0.2 MeV from the full intensity distribution are shown in
Fig.~\ref{fig:siggg}. These values are compared with those given in
Table~\ref{tab:gam} for resolved transitions in $^{90}$Zr. One sees that the
two curves have similar structures caused by the prominent peaks. However, the
curve including also the continuum part of the spectrum contains altogether a
strength that is by a factor of about 2.7 greater than the strength of the
resolved peaks only.
The full intensity distribution (resolved peaks and continuum) and the
corresponding scattering cross sections shown in Fig.~\ref{fig:siggg} contain
ground-state transitions and, in addition, branching transitions to lower-lying
excited states (inelastic transitions) as well as transitions from those states
to the ground state (cascade transitions). The different types of transitions
cannot be clearly distinguished. However, for the determination of the
photoabsorption cross section and the partial widths $\Gamma_0$ the intensities
of the ground-state transitions are needed. Therefore, contributions of
inelastic and cascade transitions have to be subtracted from the spectra. We
corrected the intensity distributions by simulating $\gamma$-ray cascades
\cite{rus07} from the levels with $J$ = 0, 1, 2 in the whole energy range
analogously to the strategy of the Monte-Carlo code DICEBOX \cite{bec98}. 
In these simulations, 1000 nuclear realizations starting from the ground state
were created with level densities derived from experiments \cite{egi05}. We
applied the statistical methods also for the low-energy part of the level
scheme instead of using experimentally known low-lying levels in $^{90}$Zr
because this would require the knowledge of the partial decay widths of all
transitions populating these fixed levels. Fluctuations of the
nearest-neighbor-spacings were taken into account according to the Wigner
distribution (see, e.g., Ref. \cite{bro81}). The partial widths of the
transitions to low-lying levels were assigned using a priori known strength
functions for $E1$, $M1$, and $E2$ transitions. Fluctuations of the partial
widths were treated the applying Porter-Thomas distribution \cite{por56}.

In the calculations, the recently published parameters for the Back-Shifted
Fermi-Gas (BSFG) model obtained from fits to experimental level densities
\cite{egi05}, $a$ = 8.95(41) MeV$^{-1}$ and $E_1$ = 1.97(30) MeV, were used.
In the individual nuclear realizations, the values of $a$ and $E_1$ were varied
within their uncertainties. As usual in the BSFG model, we assumed equal level
densities for states with positive and negative parities of the same spin
\cite{egi05}. This assumption has recently been justified by the good agreement
of level densities predicted by the BSFG model with experimental level
densities of $1^+$ states in the energy range from 5 to 10 MeV obtained from
the $^{90}$Zr($^3$He,$t$)$^{90}$Nb reaction \cite{kal06} and with experimental
level densities of $2^+$ and $2^-$ states in $^{90}$Zr studied in the
$^{90}$Zr$(e,e')$ and $^{90}$Zr$(p,p')$ reactions \cite{kal07}. The extended
analysis of the $^{90}$Zr($^3$He,$t$)$^{90}$Nb reaction in Ref.~\cite{kal07}
indicates however fluctuations of the level density of $1^+$ states in
$^{90}$Nb around the predictions of the BSFG model.

For the $E1$, $M1$, and $E2$ photon strength functions Lorentz parametrizations
\cite{axe62} were used. The parameters of the Lorentz curve for the $E1$
strength were taken from a fit to $(\gamma,n)$ data \cite{ber67} and are
consistent with the Thomas-Reiche-Kuhn sum rule (TRK)
$\frac{\pi}{2} \sigma_0 \Gamma = 60 NZ/A$ MeV mb \cite{rin80}. The parameters
for the $M1$ and $E2$ strengths were taken from global parametrizations of $M1$
spin-flip resonances and $E2$ isoscalar resonances, respectively \cite{rip06}.

\begin{figure}
\epsfig{file=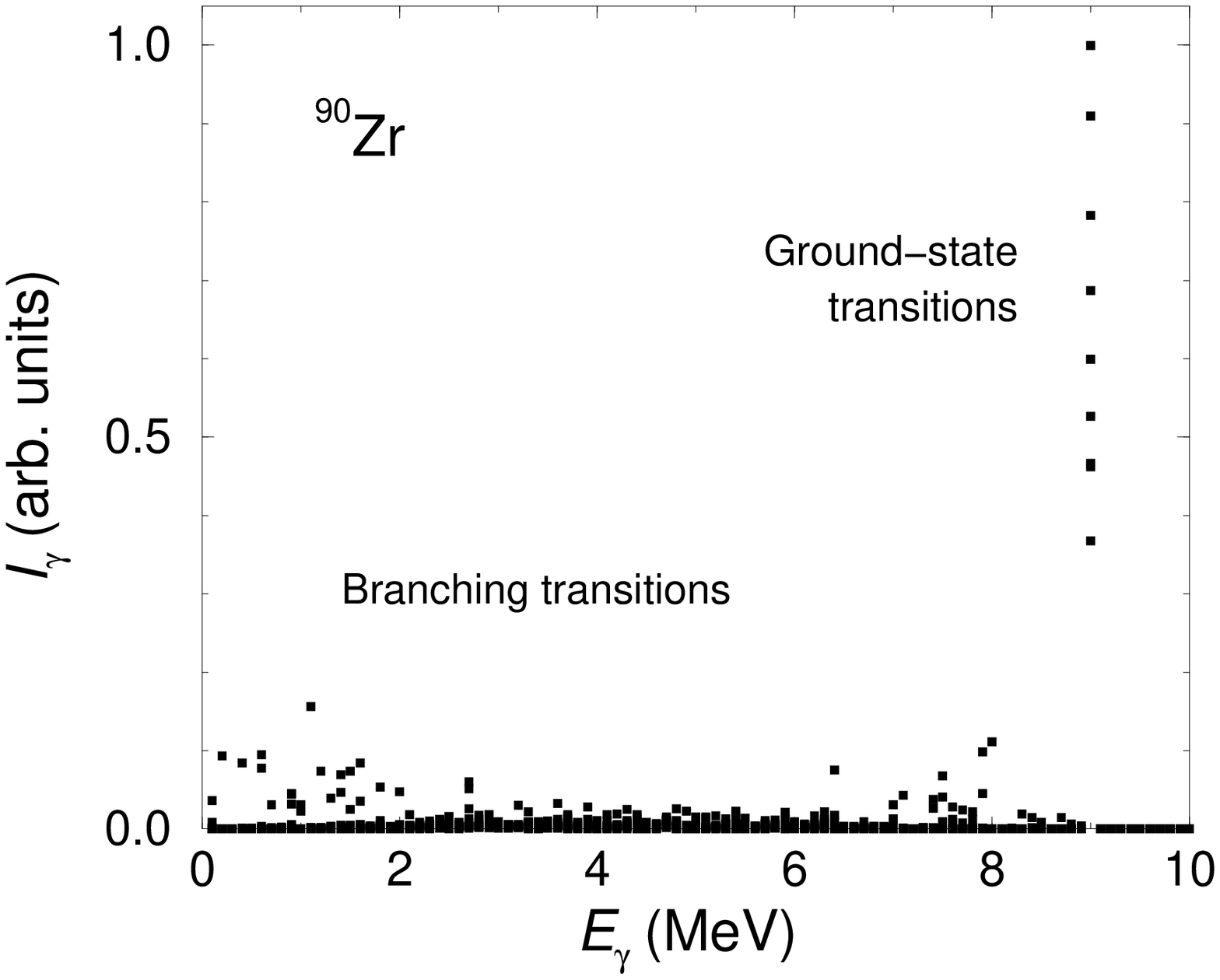,width=8cm}
\caption{\label{fig:casc} Simulated intensity distribution of transitions
depopulating levels in a 100 keV bin around 9 MeV in $^{90}$Zr. The squares
depict the intensities obtained from 10 individual nuclear realizations.}
\end{figure}

Spectra of $\gamma$-ray cascades were generated for groups of levels in
100 keV bins in each of the 1000 nuclear realizations. For illustration, the
distributions resulting from 10 individual nuclear realizations populating
levels in a 100 keV bin around 9 MeV are shown in Fig.~\ref{fig:casc}, which
reflect the influence of fluctuations of level energies and level widths.
Because in the nuclear realizations the levels were created randomly starting
from the ground state instead of starting with the known first excited state at
2.2 MeV, the distribution of the branching transitions continues to the energy
bin of the ground-state transitions.
These spectra resemble qualitatively the ones measured in an experiment on
$^{90}$Zr using tagged photons \cite{ala87}. Starting from the high-energy end
of the experimental spectrum, which contains ground-state transitions only, the
simulated intensities of the ground-state transitions were normalized to the
experimental ones in the considered bin and the intensity distribution of the
branching transitions was subtracted from the experimental spectrum. Applying
this procedure step-by-step for each energy bin moving toward the low-energy
end of the spectrum one obtains the intensity distribution of the ground-state
transitions. Simultaneously, the branching ratios $b_0^\Delta$ of the
ground-state transitions are deduced for each energy bin $\Delta$. In an
individual nuclear realization, the branching ratio $b_0^\Delta$ is calculated
as the ratio of the sum of the intensities of the ground-state transitions from
all levels in $\Delta$ to the total intensity of all transitions depopulating
those levels to any low-lying levels including the ground state 
\cite{sch07,rus08}. By dividing the summed intensities in a bin of the
experimental intensity distribution of the ground-state transitions by the
corresponding branching ratio we obtain the absorption cross section for a bin
as $\sigma_\gamma^\Delta = \sigma_{\gamma \gamma}^\Delta/b_0^\Delta$. Finally,
the absorption cross sections of each bin were obtained by averaging over the
values of the 1000 nuclear realizations. For the uncertainty of the absorption
cross section a 1$\sigma$ deviation from the mean has been taken.

\begin{figure}
\epsfig{file=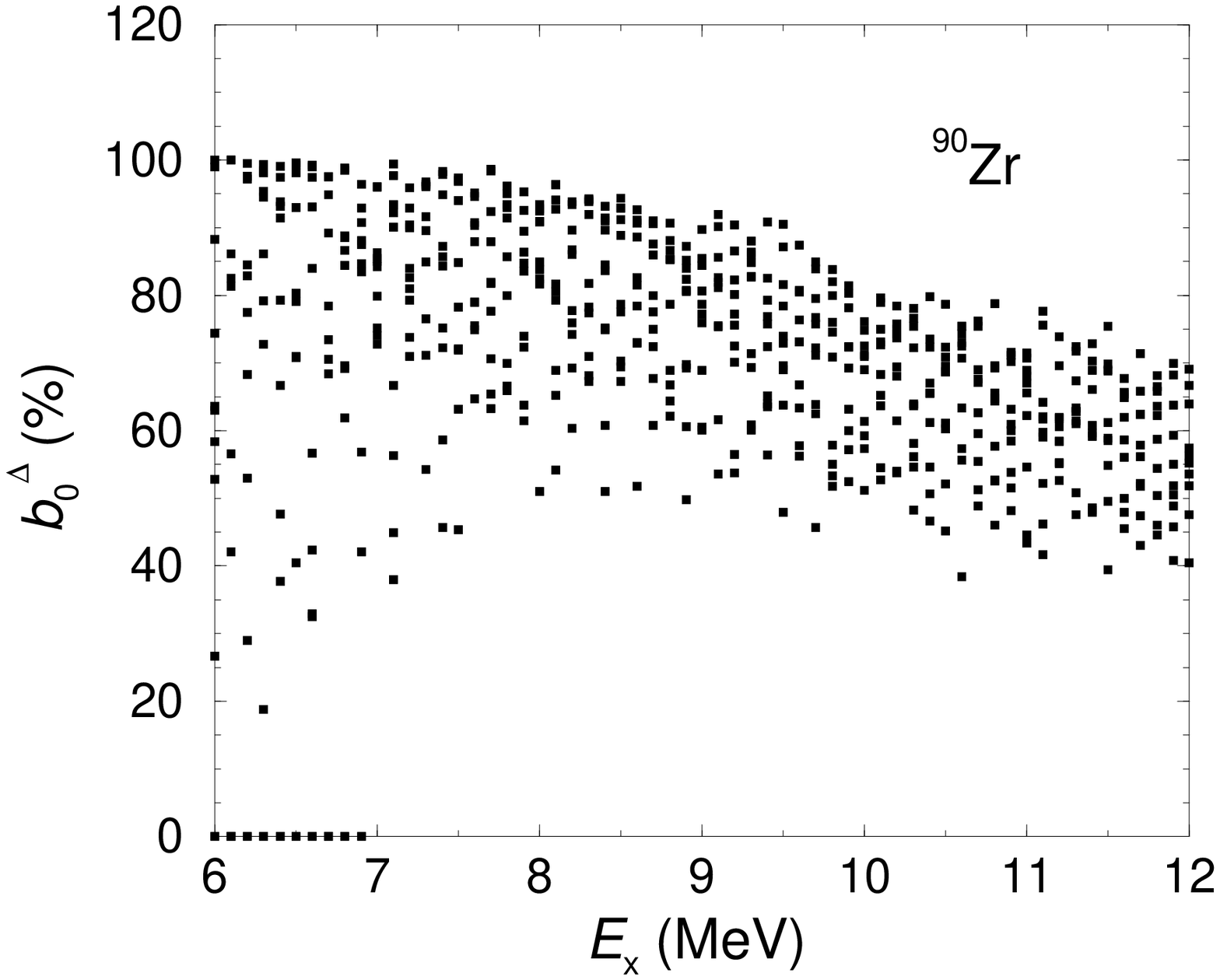,width=8cm}
\caption{\label{fig:b0} Branching ratios of ground-state transitions as
obtained from the simulations of $\gamma$-ray cascades for $^{90}$Zr. The
squares represent the values of 10 individual nuclear realizations.}
\end{figure}

To get an impression about the branching ratios, the individual values
of 10 nuclear realizations are shown in Fig.~\ref{fig:b0}. The mean branching
ratio of the 1000 realizatios decreases from about 80\% for low-lying states,
where only few possibilities for transitions to lower-lying states exist, to
about 65\% at the neutron-separation energy. Toward low energy the uncertainty
of $b_0^\Delta$ increases due to level-spacing fluctuations and the decreasing
level density. The large fluctuations below about 6 MeV make these values
useless. Note that the mean branching ratio is not representative for
transitions with large intensities like the resolved transitions given in
Table~\ref{tab:gam}. It turns out from the simulations that the branching
ratios of transitions with partial widths $\Gamma_0$ like the ones given in
Table~\ref{tab:gam} are in the order of $b_0 \approx$ 85\% to 99\%
(cf. also Ref.~\cite{sch07}).

\begin{figure}
\epsfig{file=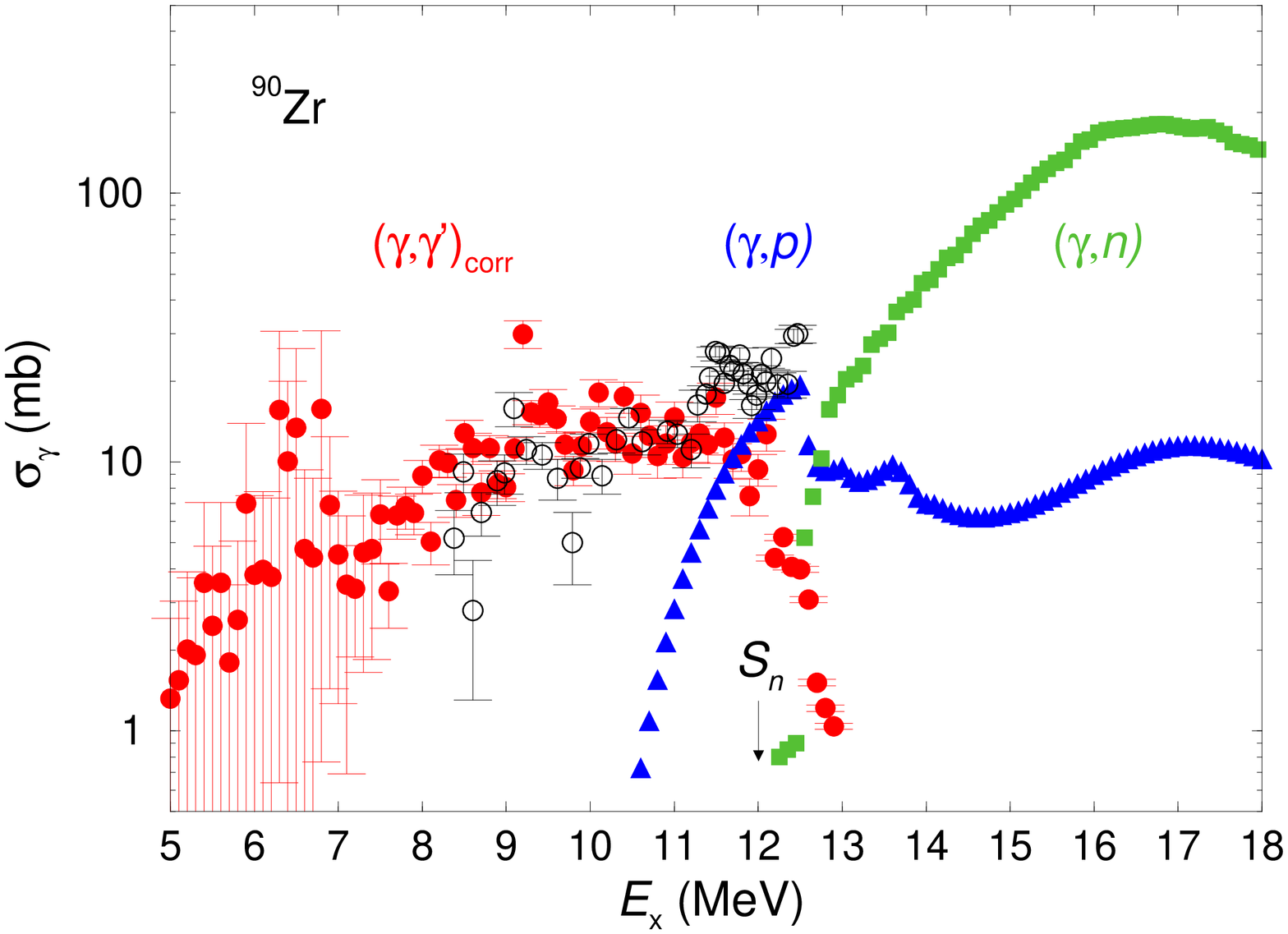,width=8cm,angle=-90}
\caption{\label{fig:sigabs}(Color online) Photoabsorption cross section deduced
from the present $(\gamma,\gamma')$ data for $^{90}$Zr after correction for
branching transitions (red circles) in comparison with data obtained from an
experiment with monoenergetic photons (open black circles) \cite{dat73},
$(\gamma,n)$ data taken from Ref.~\cite{ber67} (green boxes) and with
$(\gamma,p)$ cross sections calculated with the Talys code \cite{kon05}
(blue triangles).}
\end{figure}

The photoabsorption cross sections obtained for $^{90}$Zr are compared with the
data of a previous experiment with monoenergetic photons in the energy range
from 8.4 to 12.5 MeV \cite{axe70} in Fig.~\ref{fig:sigabs}. The data of
Ref.~\cite{axe70} contain corrections in form of additional constant partial
widths for branching and for proton emission, i.e. for the $(\gamma,p)$
channel. Cross sections obtained in $(\gamma,n)$ experiments \cite{ber67} and
cross sections calculated for the $(\gamma,p)$ reaction using the Talys code
\cite{kon05} with the Lorentz model for the $E1$ strength function
\cite{axe62} are also shown in Fig.~\ref{fig:sigabs}. 

\begin{figure}
\epsfig{file=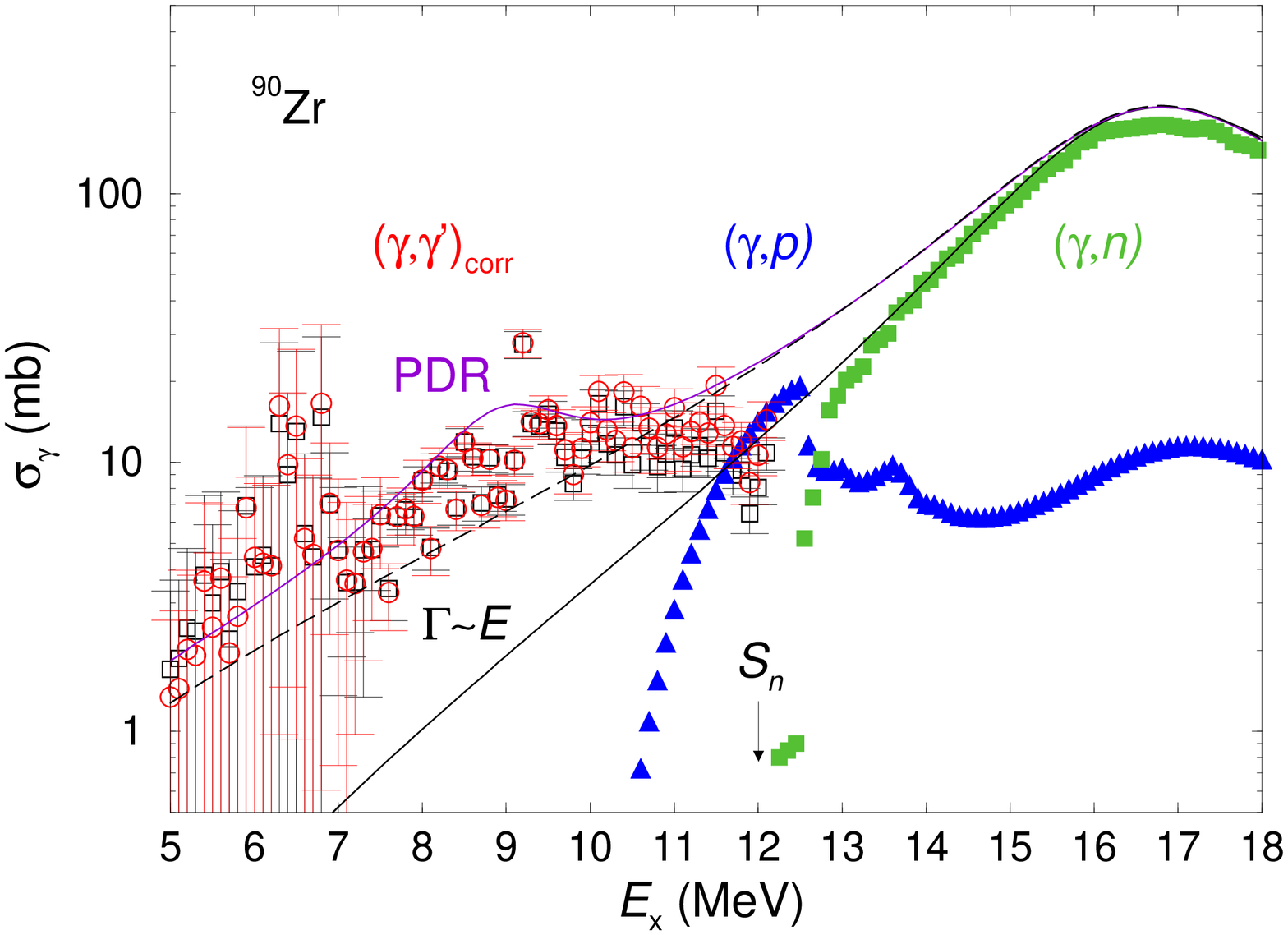,width=8cm,angle=-90}
\caption{\label{fig:D1PDR}(Color online) Photoabsorption cross sections deduced
from the present $(\gamma,\gamma')$ data for $^{90}$Zr like the red circles in
Fig.~\ref{fig:sigabs}, but with modified $E1$ strength functions. The black
squares are the results for a Lorentz curve with energy-dependent width (black
line) and the red circles are the results for the assumption of a PDR on top of
a Lorentz curve with constant width (pink line). The Lorentz curve with a 
constant width and without PDR is shown as a black dashed line.}
\end{figure}

To test the influence of the $E1$ strength function on the results of the
simulation and to check the consistency between input strength function and
deduced photoabsorption cross section we performed cascade simulations with
modified $E1$ strength functions. In one case, we applied a Lorentz curve with
an energy-dependent width $\Gamma(E) \sim E$. In another case we assumed a PDR
on top of the Lorentz curve with constant width. The PDR had a Breit-Wigner
shape with its maximum at 9 MeV and a width of 2 MeV. The cross section of the
PDR was chosen 2.2\% of the TRK (see discussion in Sec.~\ref{sec:QPMres}).
Cross sections corresponding to these input strength functions and the obtained
absorption cross sections are shown in Fig.~\ref{fig:D1PDR}. One sees that
the resulting absorption cross sections differ only slightly from each other
and also from the ones obtained by using the Lorentz curve with constant width
(cf. Fig.~\ref{fig:sigabs}) and overlap within their uncertainties. This shows
that the resulting cross sections are not very sensitive to modifications of
the input strength functions. However, the comparison of the resulting cross
sections with the input strength functions shows that there is consistency in
the case of the Lorentz curve with a PDR in addition, whereas there is a large
discrepancy in the case of the Lorentz curve with energy-dependent width.
Summarizing, the Lorentz curve with a constant width and this curve with a PDR
in addition deliver consistency between input and output strength functions and
similar results.

\begin{figure}
\epsfig{file=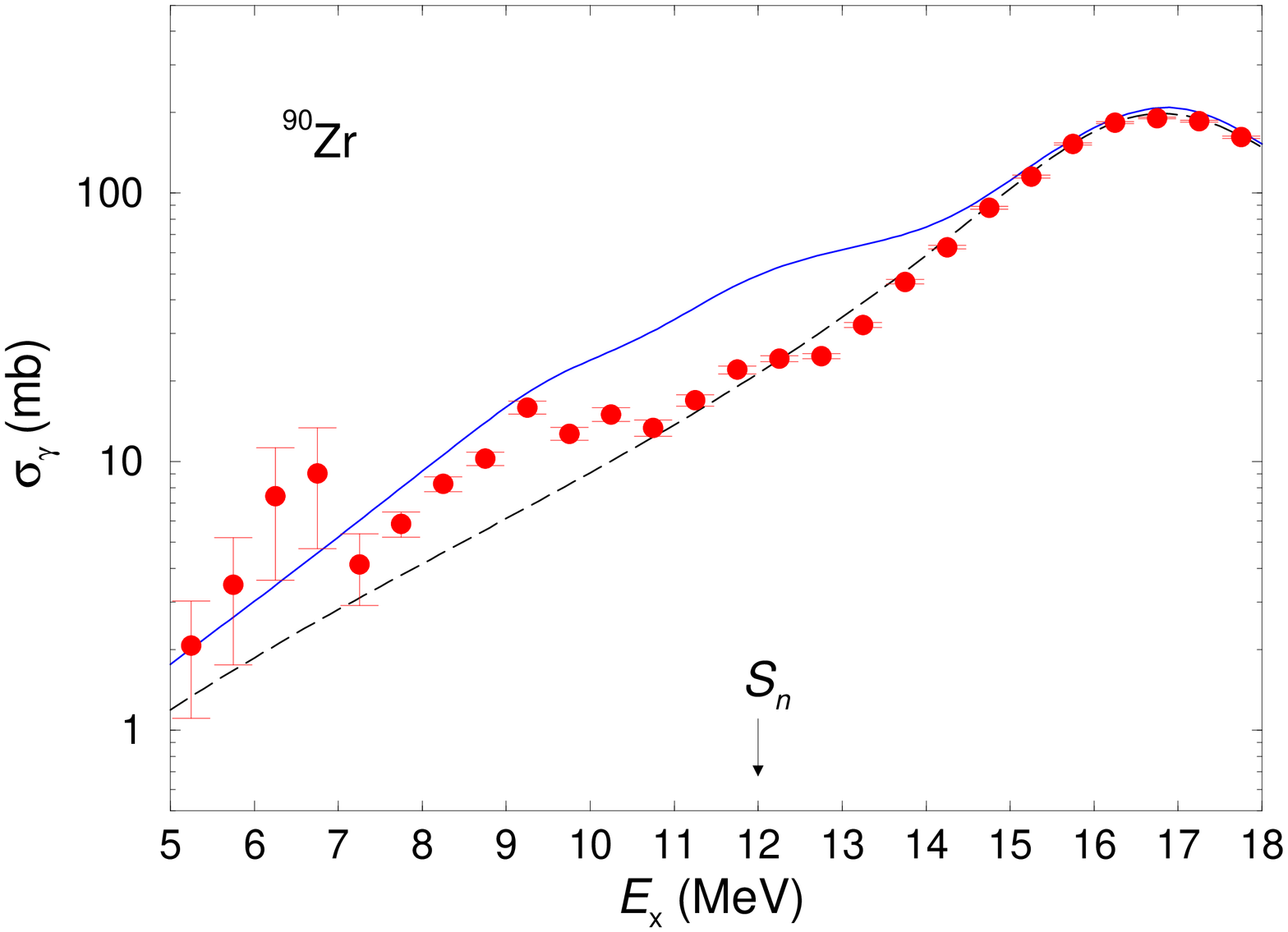,width=8cm,angle=-90}
\caption{\label{fig:sigtot90zr}(Color online) Total photoabsorption cross
section of $^{90}$Zr obtained by combining the present $(\gamma,\gamma')$ data,
the $(\gamma,n)$ data of Ref.~\cite{ber67} and the $(\gamma,p)$ calculations of
Ref.~\cite{kon05}. The data were averaged over 0.5 MeV bins to reduce
statistical fluctuations. The black dashed line represents a Lorentz curve with
the parameters taken from Ref.~\cite{ber67} as also shown in 
Fig.~\ref{fig:D1PDR}. The blue solid line is the result of QRPA calculations
discussed in Sec.~\ref{sec:QPMres}.}
\end{figure}

The total photoabsorption cross section has been deduced by combining the
present $(\gamma,\gamma')$ data with the $(\gamma,n)$ data of Ref.~\cite{ber67}
and calculated $(\gamma,p)$ cross sections \cite{kon05}. This total cross
section is compared with the Lorentz curve described above, which is currently 
used as a standard strength function for the calculation of reaction data 
\cite{rip06}, in Fig.~\ref{fig:sigtot90zr}. It can be seen that the 
experimental cross section includes extra strength with respect to the
approximation of the low-energy tail of the GDR by a Lorentz curve in the 
energy range from 6 to 11 MeV and is in better agreement with a strength
function including a PDR on top of the Lorentz curve
(cf. Fig.~\ref{fig:D1PDR}).
To investigate the nature of this extra strength we have performed model 
calculations which are discussed in the following section.

\section{Quasiparticle-phonon-model calculations for the $N$ = 50 isotones
         $^{88}$S\lowercase{r} and $^{90}$Z\lowercase{r}}
\label{sec:QPM}

The experimental data on the dipole response of $^{90}$Zr discussed in the
present work reveal a strong enhancement of the photoabsorption cross section
(see Fig. \ref{fig:sigtot90zr}) in the energy range $E_x$ = 6 -- 11 MeV and a
considerable deviation of its shape from the Lorentz-like strength function
used to adjust the GDR. Similar dipole resonance structures have been detected
also in our recent study of $^{88}$Sr \cite{sch07}. In the following we
investigate the nature of the dipole strength in the $N = 50$ isotones
$^{88}$Sr and $^{90}$Zr in the framework of the extended quasiparticle-phonon
model (QPM) \cite{Ts04,Nadia08}. The approach is based on a
Hartree-Fock-Bogoljubov (HFB) description of the ground state \cite{Ts04} by
using a phenomenological energy-density functional (EDF) \cite{Nadia08}. The
excited states are calculated within the QPM \cite{Sol76}.

\subsection{The QPM model}
\label{sec:QPMmod}

The model hamiltonian is given by:
\begin{equation}
{H=H_{MF}+H_M^{ph}+H_{SM}^{ph}+H_M^{pp}} \quad
\label{hh}
\end{equation}
and resembles in structure that of the traditional QPM model \cite{Sol76} but
in detail differs in the physical content in important aspects as discussed in
Ref.~\cite{Ts04,Nadia08}. The HFB term $H_{MF} = H_{sp} + H_{pair}$ contains
two parts. $H_{sp}$ describes the motion of protons and neutrons in a static,
spherically-symmetric mean field and $H_{pair}$ accounts for the monopole
pairing between isospin-identical particles with phenomenologically adjusted
coupling constants, fitted to data \cite{Audi95}. The mean-field hamiltonian
compromises microscopic HFB effects \cite{Hof98,Nadia08} and phenomenological
aspects like experimental separation energies and, when available, also charge
and mass radii. For numerical convenience the mean-field potential is
parametrized in terms of a superpositions of Wood-Saxon (WS) potentials with
adjustable parameters \cite{Nadia08}. 

The remaining three terms present the residual interaction
$H_{res} = H_M^{ph} + H_{SM}^{ph} + H_M^{pp}$. As typical for the QPM, we use
separable multipole-multipole $H_M^{ph}$ and spin-multipole $H_{SM}^{ph}$
interactions both of isoscalar and isovector type in the particle-hole channel
and a multipole-multipole pairing $H_M^{pp}$ in the particle-particle channel.
The isoscalar and isovector coupling constants are obtained by a fit to
energies and transition strengths of low-lying vibrational states and
high-lying GDRs \cite{Nadia08,Vdo83}.

The excited nuclear states are described by
quasiparticle-random-phase-approximation (QRPA) phonons. They are defined as
a linear combination of two-quasiparticle creation and annihilation operators:
\begin{equation}
Q^{+}_{\lambda \mu i}=\frac{1}{2}{
\sum_{jj'}{ \left(\psi_{jj'}^{\lambda i}A^+_{\lambda\mu}(jj')
-\varphi_{jj'}^{\lambda i}\widetilde{A}_{\lambda\mu}(jj')
\right)}},
\label{eq:StateOp}
\end{equation}
where $j\equiv{(nljm\tau)}$ is a single-particle proton or neutron state,
${A}^+_{\lambda \mu}$ and $\widetilde{A}_{\lambda \mu}$ are
time-forward and time-backward operators, coupling proton and neutron
two-quasiparticle creation or annihilation operators to a total
angular momentum $\lambda$ with projection $\mu$ by means of the
Clebsch-Gordan coefficients $C^{\lambda\mu}_{jmj'm'}=\left\langle
jmj'm'|\lambda\mu\right\rangle$. The time reversed operator
is defined as $\widetilde{A}_{\lambda \mu}=(-)^{\lambda
-\mu}A_{\lambda-\mu}$.

We take into account the intrinsic fermionic structure of the phonons by the
commutation relations:
\[
[Q_{\lambda \mu i}, Q^{+}_{\lambda' \mu' i'} ]
= \frac{\delta_{\lambda, \lambda'} \delta_{\mu, \mu'} \delta_{i,i'} }
{2} \sum_{jj'}
[\psi^{\lambda i}_{jj'} \psi^{\lambda i'}_{jj'} -
 \varphi^{\lambda i}_{jj'} \varphi^{\lambda i'}_{jj'}]
 \]
 \[
 - \sum_{\scriptstyle jj'j_{2} \atop \scriptstyle m m'm_{2}}
\alpha^{+}_{jm} \alpha_{j'm'}
 \times
\left \{
\psi^{\lambda i}_{j'j_{2}} \psi^{\lambda'i'}_{jj_{2}}
C_{j'm' j_2m_2}^{\lambda \mu}
C_{j m  j_2m_2}^{\lambda' \mu'}
\right.
\]
\begin{equation}
\left.
- (-)^{\lambda + \lambda'+\mu + \mu'}
\varphi^{\lambda i}_{jj_{2}} \varphi^{\lambda'i'}_{j'j_{2}}
C_{j m  j_2m_2}^{\lambda -\mu}
C_{j'm' j_2m_2}^{\lambda' -\mu'}.
\right \} ~
\label{comut}
\end{equation}
The QRPA phonons obey the equation of motion
\begin{equation}\label{eq:EoM}
\left[H,Q^+_i\right]=E_i Q^+_i,
\end{equation}
which solves the eigenvalue problem, i.e. it gives the excitation energies
$E_i$ and the time-forward and time-backward amplitudes
\cite{Sol76} $\psi_{j_1j_2}^{\lambda i}$ and $\varphi_{j_1j_2}^{\lambda i}$,
respectively.

In terms of the QRPA phonons the model hamiltonian can be expressed as
\begin{eqnarray}
 H=H_{ph} + H_{qph}=
\sum_{\lambda \mu i}\omega _{\lambda i}Q_{\lambda \mu i}^{+}Q_{\lambda
\mu i}^{}
\label{hp}
\end{eqnarray}
\begin{eqnarray}
+\frac{1}{2}\sum_{{\lambda _1\lambda _2\lambda _3}{{
{i_1i_2i_3}{\mu _1\mu _2\mu _3}}}}C_{\lambda _1\mu _1\lambda _2\mu
_2}^{\lambda _3{-\mu _3}}
U_{\lambda _1i_1}^{\lambda _2i_2}({\lambda _3i_3})
\nonumber
\end{eqnarray}
\begin{eqnarray}
[Q_{\lambda _1\mu
_1i_1}^{+}Q_{\lambda _2\mu _2i_2}^{+}Q_{\lambda _3-\mu
_3i_3}^{}+h.c.].
\nonumber
\end{eqnarray}
The first part in the Eq. (\ref{hp}) refers to the harmonic part of nuclear
vibrations, while the second one is accounting for the interaction between
quasiparticles and phonons. The latter allows us to go beyond the ideal
harmonic picture by including effects due to the non-bosonic features of the
nuclear excitations.

The model hamiltonian is diagonalized assuming a spherical $0^+$ ground state
which leads to an orthonormal set of wave functions with good total angular
momentum $JM$. For even-even nuclei we extend the approach to anharmonic
effects by considering wave functions including a mixture of one-, two-, and
three-phonon components \cite{Gri94}: 
\begin{equation}
\Psi_{\nu} (JM) =
\left\{ \sum_i R_i(J\nu) Q^{+}_{JMi}
\right.
\label{wf}
\end{equation}
\[
\left.
+ \sum_{\scriptstyle \lambda_1 i_1 \atop \scriptstyle \lambda_2 i_2}
P_{\lambda_2 i_2}^{\lambda_1 i_1}(J \nu)
\left[ Q^{+}_{\lambda_1 \mu_1 i_1} \times Q^{+}_{\lambda_2 \mu_2 i_2}
\right]_{JM}
{+ \sum_{_{ \lambda_1 i_1 \lambda_2 i_2 \atop
 \lambda_3 i_3 I}}}
\right.
\]
\[
\left.
{T_{\lambda_3 i_3}^{\lambda_1 i_1 \lambda_2 i_2I}(J\nu )
\left[ \left[ Q^{+}_{\lambda_1 \mu_1 i_1} \otimes Q^{+}_{\lambda_2 \mu_2
i_2} \right]_{IK}
\otimes Q^{+}_{\lambda_3 \mu_3 i_3}\right]}_{JM}\right\}\Psi_0,
\]
where $\nu$ labels the number of the  excited states. The energies and the
phonon-mixing amplitudes $R$, $P$, and $T$ are determined by solving the
extended multi-phonon equations of motions.

As discussed in Refs.~\cite{Rye02,Nadia08} we obtain important additional
information from the analysis of the spatial structure of the nuclear response
on an external electromagnetic field. That information is accessible by
considering the one-body transition densities $\delta\rho (\vec{r})$, which are
the non-diagonal elements of the nuclear one-body density matrix. The
transition densities are obtained by the matrix elements between the ground
state $|\Psi_i\rangle=|J_iM_i\rangle$ and the excited states
$|\Psi_f\rangle=|J_fM_f\rangle$. We identify in QPM
$|J_iM_i\rangle\equiv |0\rangle$ with phonon vacuum and obtain the excited
states by means of the QRPA state operator, Eq. (\ref{eq:StateOp}),
$|J_fM_f\rangle\equiv Q^+_{\lambda \mu i}|0\rangle$.
The analytical procedure for the calculation of the transition densities is
presented in details in Ref.~\cite{Nadia08}.

\subsection{Calculation of dipole excitations in the $N$ = 50 isotones}
\label{sec:QPMres}

\begin{figure}
\epsfig{file=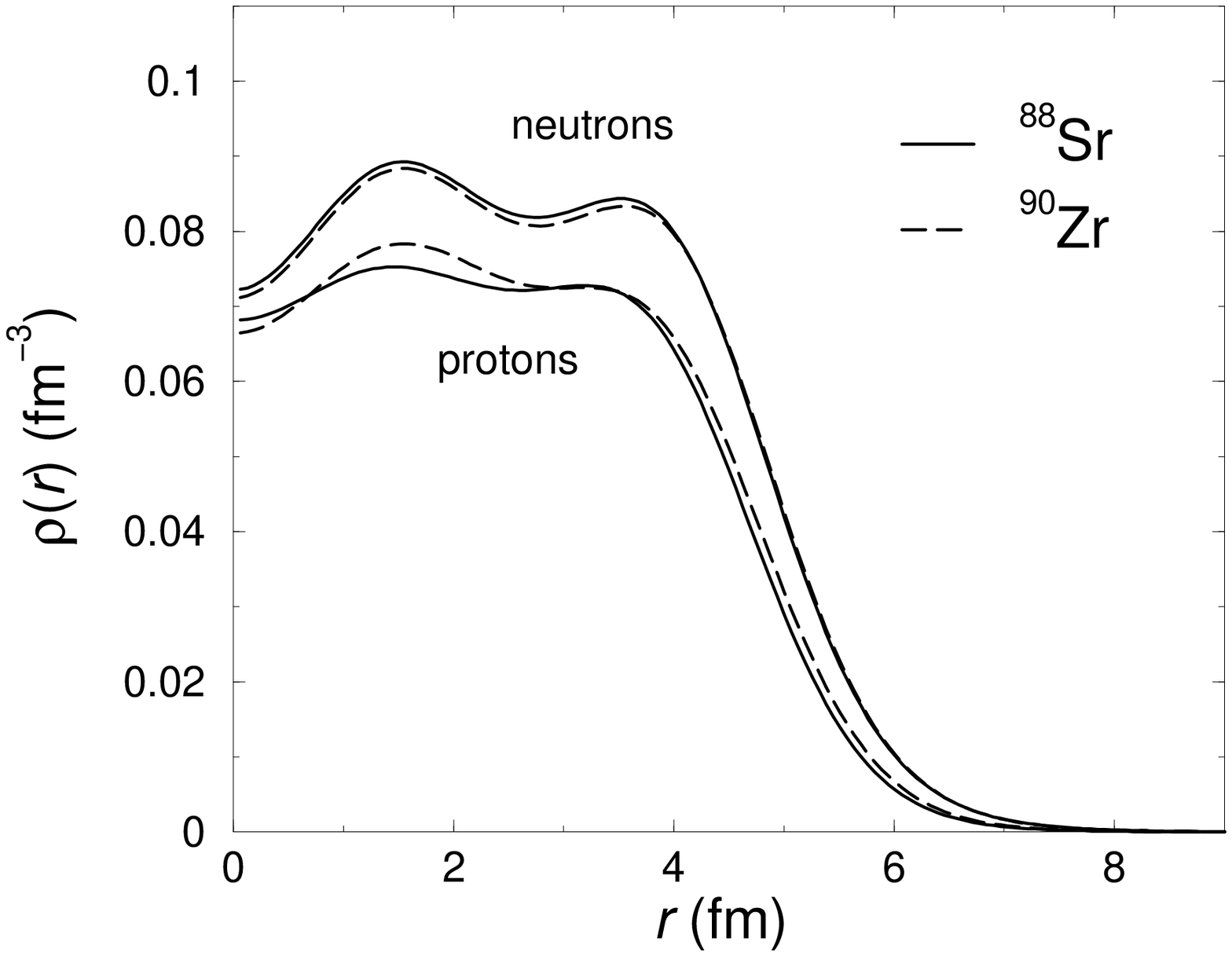,width=8cm}
\caption{\label{FIG1-QPM}Proton and neutron ground-state densities of the
$N = 50$ isotones $^{88}$Sr and $^{90}$Zr.}
\end{figure}

In Ref.~\cite{Nadia08}, a close relationship between skins in the ground-state
matter-density distributions and the low-energy dipole response was pointed
out. It was shown, that the skin thickness, defined by
\begin{equation}\label{eq:skin}
\delta r=\sqrt{<r^2_n>}-\sqrt{<r^2_p>}
\end{equation} 
is indeed directly connected to the non-energy-weighted dipole sum rule, thus
being complementary to the well established  relation of the GDR strength and
the energy-weighted Thomas-Reiche-Kuhn dipole sum rule. From the calculated
neutron and proton ground state densities displayed in Fig.~\ref{FIG1-QPM} it
is seen that the two $N = 50$ isotones considered here, $^{88}$Sr and
$^{90}$Zr, show a neutron skin. The total binding energies and radii are given
in Table~\ref{tab:HFB}. Of special importance for our investigations are the
surface regions, where the formation of a skin takes place. For the
investigated $^{88}$Sr and $^{90}$Zr nuclei, the neutron skin decreases with
decreasing $N/Z$, i.e. when going from $^{88}$Sr to $^{90}$Zr.

\begin{table}
\caption{\label{tab:HFB}HFB results obtained with the phenomenological EDF for
total binding energy per nucleon, the measured binding energy \cite{Audi95},
proton and neutron root-mean-square radii, and neutron skin thickness,
Eq.~(\ref{eq:skin}), for the two $N$ = 50 isotones.}
\begin{ruledtabular}
\begin{tabular}{cccccc} 
 & $\frac{B(A)}{A}_{\mathrm exp}$ (MeV) & 
   $\frac{B(A)}{A}_{\mathrm the}$ (MeV) &
   $\sqrt{r^2_p}$ (fm) & $\sqrt{r^2_n}$ (fm) & $\delta r$ (fm) \\
\hline
$^{88}$Sr & --8.733 & --8.766 & 4.185 & 4.291 & 0.106 \\
$^{90}$Zr & --8.710 & --8.692 & 4.229 & 4.303 & 0.074 \\
\end{tabular}
\end{ruledtabular}
\end{table}

\begin{figure}
\epsfig{file=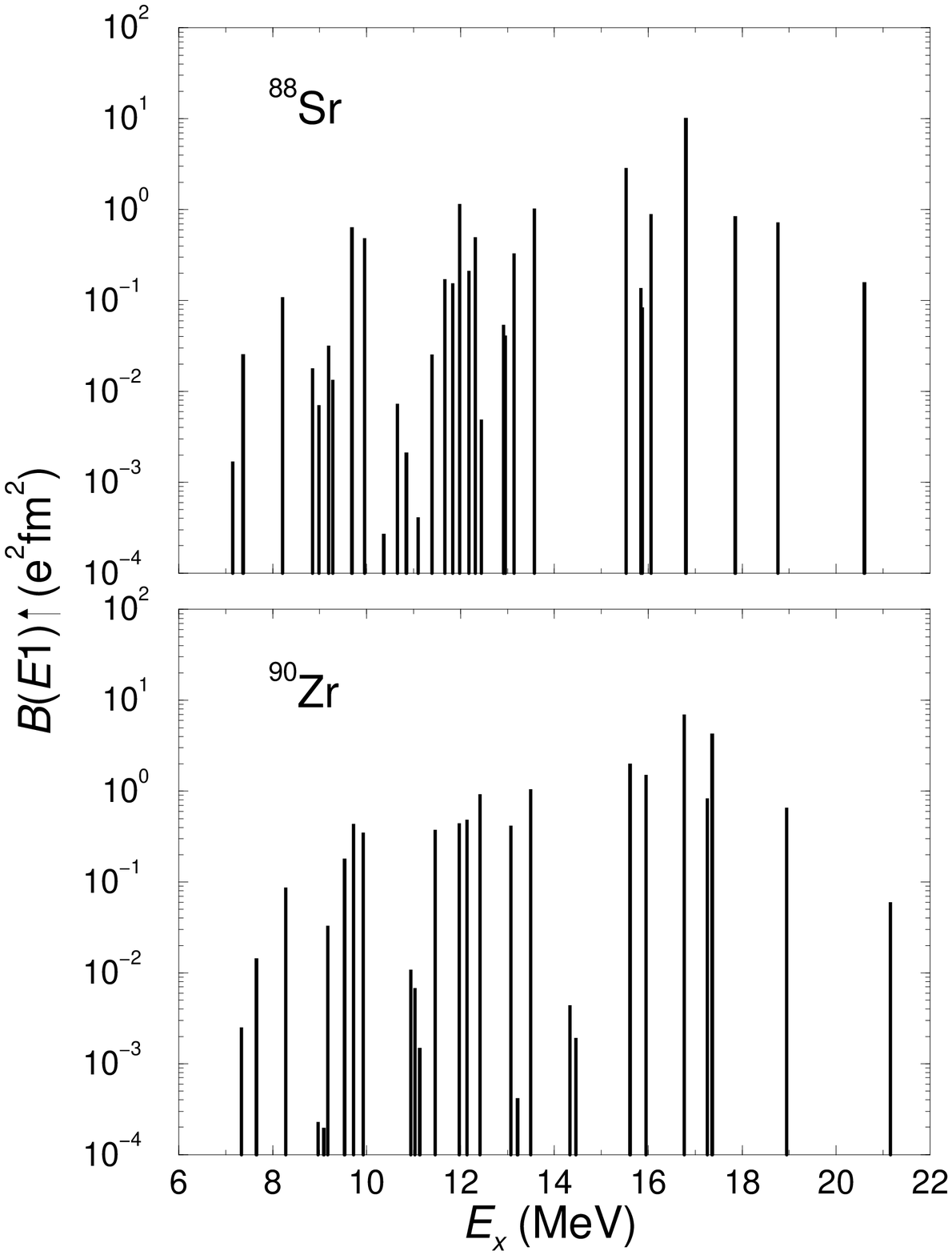,width=8cm}
\caption{\label{fig2a:theo}Calculated dipole one-phonon transition strengths
up to $E_x$ = 22 MeV for the isotones $^{88}$Sr and $^{90}$Zr.}
\end{figure}

\begin{figure}
\epsfig{file=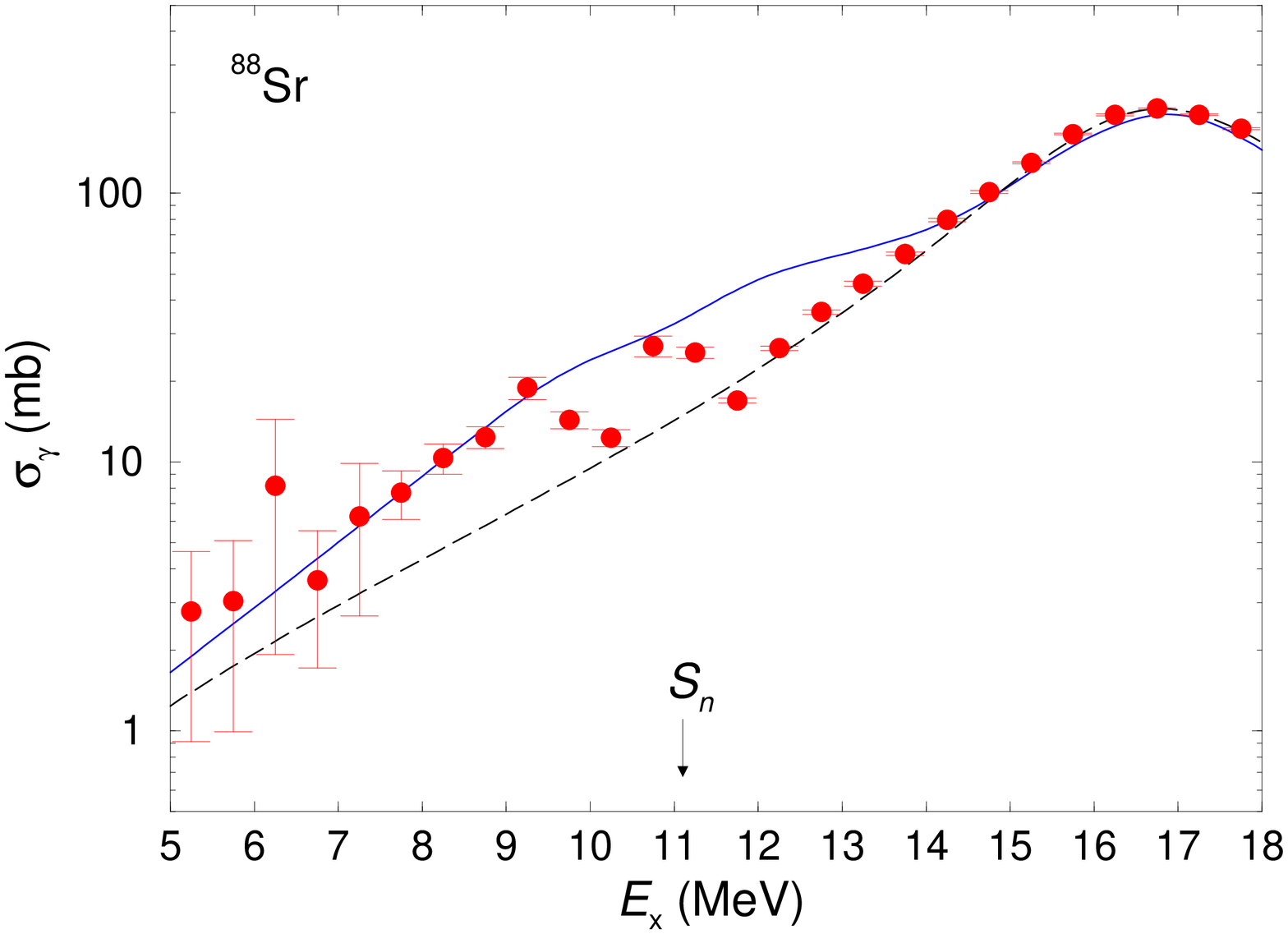,width=8cm,angle=-90}
\caption{\label{fig:sigtot88sr}(Color online) Total photoabsorption cross
section of $^{88}$Sr as taken from Ref.~\cite{sch07}. The black dashed line
represents a Lorentz distribution with the parameters given in
Ref.~\cite{sch07}. The blue solid line is the result of the present QRPA
calculations.}
\end{figure}

The QRPA calculations in these nuclei result in a sequence of low-lying
one-phonon dipole states of almost pure neutron structure, located below the
particle threshold. The proton contribution in the state vectors is less than
5\%. These states have been related to the oscillations of weakly bound
neutrons from the surface region, known as neutron PDR
(see Refs.~\cite{Rye02,Adri05,Ts04,Nadia08,Volz06} and Refs. therein). The
calculations of the dipole response in $N$ = 50 up to 22 MeV are presented in
Fig.~\ref{fig2a:theo}. The $E1$ transition matrix elements were calculated with
recoil-corrected effective charges $q_{n} = -Z/A$ for neutrons and
$q_{p} = N/A$ for protons \cite{Nadia08}, respectively. To compare the
predicted dipole strengths with the experimental values we calculated
the integrated absorption cross sections according to  
$\int \sigma_\gamma ~dE = 4.03 E_x B(E1,0^+\rightarrow 1^-)$ MeV mb with
$E_x$ in MeV and $B(E1)$ in $e^2$fm$^2$ \cite{Bohr}. When comparing the QRPA
results with experimental cross sections one has the general problem of
comparing discrete values (cf. Fig.~\ref{fig2a:theo}) with gradually changing
averages of energy bins (cf. Fig.~\ref{fig:sigtot90zr}). Therefore, the QRPA
values are usually folded with Lorentz curves that can be considered as a
simulation of the damping of the GDR due to higher-order configurations
neglected in QRPA. In the present case, we folded the discrete QRPA cross
sections with Lorentz curves of 3.0 MeV width, which is a typical value for the
damping width. The results are compared with the experimental cross sections
in Fig.~\ref{fig:sigtot90zr} for $^{90}$Zr and in Fig.~\ref{fig:sigtot88sr}
for $^{88}$Sr. The QRPA calculations predict extra strength with respect to the
Lorentz curve in qualitative agreement with the experimental findings. However,
the experimental cross sections show a clear separation of the PDR and the GDR
regions around 12 MeV, whereas the calculations seem to predict an excess of
strength up to about 14 MeV.

\begin{figure}
\epsfig{file=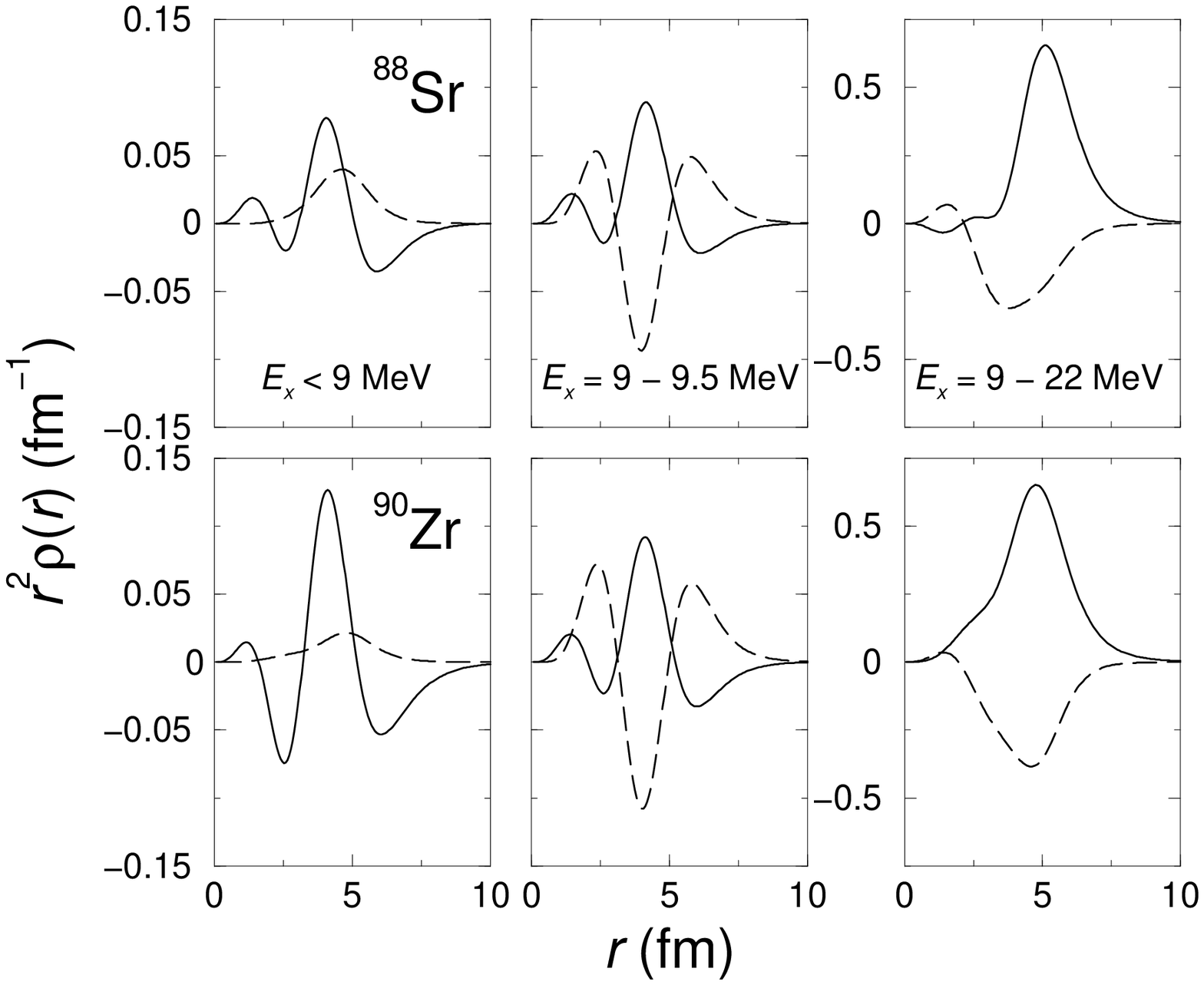,width=8cm}
\caption{\label{fig:fig2b:theo}Proton (solid lines) and neutron (dashed lines)
transition densities in the $N$ = 50 nuclei $^{88}$Sr and $^{90}$Zr.}
\label{fig2b:theo}
\end{figure}

The nature of the dipole strength can be examined by analyzing dipole
transition densities that act as a clear indicator of the PDR and allow us to
distinguish between the different types of the dipole excitations
\cite{Nadia08}. The QRPA proton and neutron dipole transition densities for
$^{88}$Sr and $^{90}$Zr were summed over selected energy regions and are shown
in Fig.~\ref{fig2b:theo}. They can be directly related to the spectrum of
1$^{-}$ states in Fig.~\ref{fig2a:theo}. In the region of $E_x <$ 9 MeV we
find an in-phase oscillation of protons and neutrons in the nuclear interior,
while at the surface only neutrons contribute. Because these features are
characteristic for dipole skin modes, we identify these states as part of the
PDR. The states in the region of $E_x$ from 9 MeV to 9.5 MeV carry a different
signature and their contribution to the total proton and neutron transition
densities in the energy region $E_x <$ 9.5 MeV changes completely the behavior
of the proton/neutron oscillations related to the PDR. At these higher energies
the protons and neutrons start to move out of phase being compatible with the
low-energy part of the GDR. At $E_x$ from 9 to 22 MeV a strong isovector
oscillation corresponding to the excitation of the GDR is observed.

In the case of $N = 50$ isotones the neutron number is fixed and only the
proton number changes. This affects the thickness of the neutron skin as well
as is seen in Fig.~\ref{FIG1-QPM} and Table~\ref{tab:HFB}. Correspondingly, the
total PDR strength decreases with increasing proton number. According to the
criteria discussed above we consider the states in the energy region
$E_x <$ 9 MeV as a part of the PDR. In this energy interval, the QRPA
calculations predict a total PDR strength of 
$\sum B(E1)$$\uparrow = 0.107~e^2$fm$^2$ in $^{90}$Zr and
$\sum B(E1)$$\uparrow = 0.170~e^2$fm$^2$ in $^{88}$Sr, respectively.
This accounts for 0.27\% and 0.44\% of the TRK in $^{90}$Zr and $^{88}$Sr,
respectively, and compares with experimental values of 0.55(2)\% and 0.74(2)\%
derived from resolved peaks in $^{90}$Zr and $^{88}$Sr, respectively, whereas
the values deduced from the corrected full intensities including also the
continuum part are remarkably greater and amount to 2.2(4)\% and 2.4(6)\% in
$^{90}$Zr and $^{88}$Sr, respectively.

Finally, we note that the results are in agreement with the established
connection between the calculated total PDR strength and the nuclear skin
thickness, Eq.~(\ref{eq:skin}), found in our previous investigations in the
$N$ = 82 isotones \cite{Volz06} and the $^{100-132}$Sn isotopes \cite{Nadia08}
nuclei.

In the multi-phonon QPM calculations the structure of the excited states is
described by wave functions as defined in Eq. (\ref{wf}). We now investigate
multi-phonon effects using a model space with up to three-phonon components,
built from a basis of QRPA states with $J^\pi$ = 1$^+$, 1$^-$, 2$^+$, 3$^-$,
4$^+$, 5$^-$, 6$^+$, 7$^-$, and 8$^+$. Because the one-phonon configurations up
to $E_x$ = 22 MeV are considered the core polarization contributions to the
transitions of the low-lying 1$^-$ states are taken into account explicitly.
Hence, we do not need to introduce dynamical effective charges. Our model space
includes one-, two- and three-phonon configurations (in total 200) with
energies up to $E_x$ = 12 MeV. 

\begin{figure}
\epsfig{file=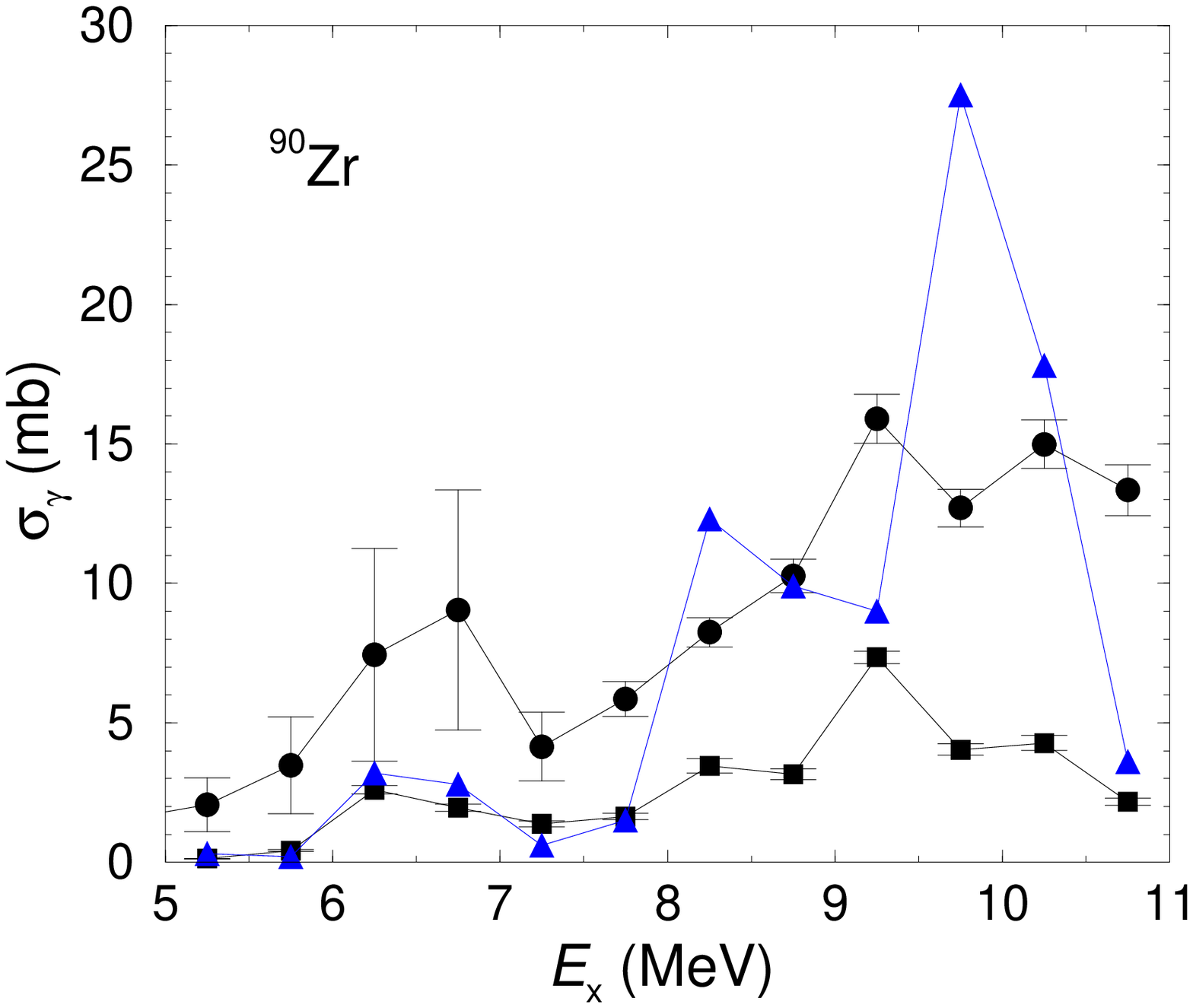,width=8cm}
\caption{\label{fig3:theo}(Color online) Experimental cross sections derived
from resolved peaks (squares), from the full intensity corrected for inelastic
transitions and branching ratios (circles), and cross sections predicted by QPM
calculations (blue triangles) for $^{90}$Zr, averaged over energy bins of 0.5
MeV. Lines are drawn to guide the eye.}
\end{figure}

The experimental absorption cross sections in $^{90}$Zr in the range from 5 to
11 MeV are compared with the corresponding QPM results in Fig. \ref{fig3:theo}.
The strength in the region $E_x$ = 6 -- 7 MeV is related to the PDR as
identified from the structure of the involved 1$^-$ states. The contribution
of the GDR phonons becomes significant at energies $E_x >$ 7.5 MeV where a
coupling between PDR, GDR, and multiphonon states reflects the properties of
the dipole excitations at higher energies. At low energy, the calculated values
underestimate the experimental values derived from the full intensity and are
close to the values deduced from resolved peaks only, whereas at high energy
they are in better agreement with the experimental values including the full
intensity. The calculated EWSR from 5 to 9 MeV of 1.3\% of the TRK
underestimates the experimental value of 2.2(4)\%. However, increasing the
energy range up to 11 MeV the theoretical value of 3.3\% is much closer to the
experimental result of 4.0(3)\%, indicating a slightly different distribution
of the QPM strength with respect to excitation energy.

\section{Summary}

The dipole-strength distribution in $^{90}$Zr up to the neutron-separation
energy has been studied in photon-scattering experiments at the ELBE
accelerator using various electron energies. Ground-state transitions have
been identified by comparing the transitions observed at different electron
energies. We identified about 190 levels with a total dipole strength of about
180 meV/MeV$^3$. Spin $J$ = 1 could be deduced from the angular correlations
of the ground-state transitions for about 140 levels including a dipole
strength of about 155 meV/MeV$^3$.

The intensity distribution obtained from the measured spectra after a
correction for detector response and a subtraction of atomic background in the
target contains a continuum part in addition to the resolved peaks. It turns
out that the dipole strength in the resolved peaks amounts to about 27\% of the
total dipole strength while the continuum contains about 73\%.

An assignment of inelastic transitions to particular levels and, thus, the
determination of branching ratios was in general not possible. To get
information about the intensities of inelastic transitions to low-lying levels
we have applied statistical methods. By means of simulations of $\gamma$-ray
cascades intensities of branching transitions could be estimated and subtracted
from the experimental intensity distribution and the intensities of
ground-state transitions could be corrected in average for their branching
ratios.

A comparison of the photoabsorption cross section obtained in this way from the
present $(\gamma,\gamma')$ experiments with $(\gamma,n)$ data shows a smooth
connection of the data of the two different experiments and gives new
information about the extension of the dipole-strength function toward
energies around and below the threshold of the $(\gamma,n)$ reaction. In
comparison with a straightforward approximation of the GDR by a Lorentz curve
one observes extra $E1$ strength in the energy range from 6 to 11 MeV which is
mainly concentrated in strong peaks.

QPM calculations in $^{88}$Sr and $^{90}$Zr predict low-energy dipole strength
in the energy region from 6 to 10 MeV.  The states at about 6 to 7.5 MeV have a
special character. Their structure is dominated by neutron components and their
transition strength is directly related to the size of a neutron skin. Their
generic character is further confirmed by the shape and structure of the
related transition densities, showing that these PDR modes are clearly
distinguishable from the GDR. The dipole states, seen as a rather fragmented
ensemble at $E_x >$ 7.5 MeV mix strongly with the low-energy tail of the GDR
starting to appear in the same region. The complicated structure of these
states and the high level densities imposes considerable difficulties for a
reliable description of the fragmentation pattern.

The present analysis shows that standard strength functions currently used
for the calculation of reaction data do not describe the dipole-strength
distribution below the $(\gamma,n)$ threshold correctly and should be improved
by taking into account the observed extra strength.

\section{Acknowlegdments}

We thank the staff of the ELBE accelerator for their cooperation during the
experiments and we thank A. Hartmann and W. Schulze for their technical
assistance. Valuable discussions with F. D\"onau and S. Frauendorf are
gratefully acknowledged. This work was supported by the
Deutsche Forschungsgemeinschaft under contracts Do466/1-2 and Le439/4-3.

\end{document}